\documentclass[preprint,12pt]{elsarticle}
\newcommand*{\halffactor}{0.5}	

\usepackage{hyperref}
\usepackage[table]{xcolor}
\usepackage{enumerate}
\usepackage{amsmath}
\usepackage{float}
\usepackage{amssymb}
\usepackage{graphicx}
\usepackage[latin1]{inputenc}
\usepackage{amsfonts}
\usepackage{hyperref}
\usepackage{framed}

\usepackage[multidot]{grffile}

\journal{Ocean Engineering}

\begin{document}

\begin{frontmatter}



\title{Analytical Prediction of Reflection Coefficients for Wave Absorbing Layers in Flow Simulations of Regular Free-Surface Waves}


\author{Robinson Peri\'c\footnote{Corresponding author. E-mail address: robinson.peric@tuhh.de}, Moustafa Abdel-Maksoud}
\address{Institute for Fluid Dynamics and Ship Theory, Hamburg University of Technology (TUHH), Schwarzenbergstrasse 95 C, 21073 Hamburg, Germany}

\begin{abstract}
Undesired wave reflections, which occur at domain boundaries in flow simulations with free-surface waves, can be minimized by applying source terms in the vicinity of the boundary to damp the waves.
Examples of such approaches are absorbing layers, damping zones, forcing zones, relaxation zones and sponge layers.
A problem with these approaches is that the effectivity of the wave damping depends on the parameters in the source term functions, which are case-dependent and must be adjusted to the wave.  
The present paper presents a  theory which analytically predicts the reflection coefficients and which can be used to optimally select the source term parameters before running the simulation. The theory is given in a general form so that it is applicable to many existing implementations.
It is validated against results from finite-volume-based flow simulations of regular free-surface waves and found to be of satisfactory accuracy for practical purposes. 
\end{abstract}


\end{frontmatter}

\definecolor{RED}{HTML}{FF0000}
\definecolor{GREEN}{HTML}{00FF00}
\definecolor{BLUE}{HTML}{0000FF}
\definecolor{PINK}{HTML}{FF0DFF} 
\definecolor{TURQUOISE}{HTML}{0BFFFF} 
\definecolor{YELLOW}{HTML}{FFFF02} 
\definecolor{BLACK}{HTML}{000000} 
\definecolor{ORANGE}{HTML}{FF4D01} 
\definecolor{GREY}{HTML}{818181} 

\definecolor{g}{HTML}{039300}

\newcommand*{\factor}{1.0}

\section{Introduction}
\label{SECintro}
In flow simulations, it is usually desired to choose the computational domain as small as possible to reduce the computational effort. However, when simulating free-surface wave propagation, undesired wave reflections at the domain boundaries must be minimized. If this is not achieved, it can lead to large errors in the results. 
For practical purposes, it is desired to be able to estimate the amount of undesired wave reflection before running the simulations.  Of the various techniques for reducing undesired reflections (see Peri\'c and Abdel-Maksoud, 2016), this paper is concerned with the ones that apply source terms to the governing equations in a zone adjacent to the corresponding domain boundaries. Such approaches have been presented under many different names, such as absorbing layers (e.g. Wei et al., 1999), damping zones (e.g. Park et al., 1999; Peri\'c and Abdel-Maksoud, 2016; Jose et al., 2017), dissipation zones (Park et al., 1993), numerical beach (e.g. Cl\'{e}ment, 1996), sponge layers (e.g. Israeli and Orszag, 1981; Larsen and Dancy, 1983; Brorsen and Helm-Petersen, 1999; Ha et al. 2013; Choi and Yoon, 2009; Zhang et al., 2014; Hu et al., 2015), Euler overlay method (e.g. Kim et al., 2012; Kim et al., 2013), forcing zones (Peri\'c, 2015; Siemens STAR-CCM+ manual version 11.06), and coupling or relaxation zones (e.g. Jacobsen et al., 2012; W\"ockner-Kluwe, 2013; Schmitt and Elsaesser, 2015; Jasak et al., 2015; Vuk\v{c}evi\'{c} et al., 2016). 

The general principle behind all these approaches is that they apply source terms to one, to several, or to all of the governing equations, with the intention of gradually forcing the solution towards some reference solution within a zone (layer) attached to the domain boundary. This damps waves which travel into the layer, but it can also be used to generate waves or to couple different flow solvers (e.g. a viscous solver for the near-field and an inviscid solver for the far-field). 

A possible distinction could be that terms like  \textit{forcing zone} and \textit{relaxation zone} are more general, while others are more specific; for example, the \textit{Euler overlay method} forces the flow towards the analytical solution of an undisturbed wave. \textit{Damping zones}, \textit{absorbing layers} and \textit{sponge layers} often apply source terms only in a single governing equation, with the forcing term formulated so that it can be interpreted as a damping term. 
Yet in several cases there seems to be no clear distinction and some of the names are used synonymously.

Thus in the following, the term \textit{forcing zone} will be used to highlight that the results in this work are applicable to all of the above approaches. 
In Sect. \ref{SECforc}, a general formulation for forcing zones is presented, by which the results in this work can be applied to the other approaches.

The main goal with forcing-zone-type approaches is  to achieve reliable minimization of undesired wave reflections. However, the source term function contains tunable parameters which must be adjusted for every simulation (Mani, 2012; Peri\'c and Abdel-Maksoud, 2016). It was argued that these "parameters and profiles [...] can only be determined by trail and error" (Colonius, 2004, p. 337), which at the time of writing is still common practice in flow simulations with ocean waves (Peri\'c and Abdel-Maksoud, 2016). So far, no theoretical foundation has been presented which guides how to choose these parameters and predicts reflection coefficients (Colonius, 2004; Hu, 2008; Modave et al., 2010; Mani, 2012). This is especially problematic since in industrial practice reflection coefficients are seldom evaluated; instead, mostly the default coefficients are used, which can lead to significant errors (Peri\'c and Abdel-Maksoud, 2016). Thus prediction and minimization of undesired wave reflections is a key problem of great practical importance.

The present work  aims to solve this problem by presenting an analytical approach to predict the reflection coefficients for forcing-zone-type approaches. The theory can be used to tune the forcing zone parameters before running the simulation. Thus undesired wave reflections can be reliably minimized. The theory is implemented in a computer program, which is made publicly available as free software.\footnote{The source code and manual can be downloaded from: \url{https://github.com/wave-absorbing-layers/absorbing-layer-for-free-surface-waves}}

The theory is derived in Sect. \ref{SECtheory} and validated in Sect. \ref{SECresdisc} using results from finite-volume-based flow simulations; the simulation setup is given in Sect. \ref{SECsetup}. The theory is shown to predict optimum forcing zone parameters and corresponding reflection coefficients for forcing of $x$-momentum, or $z$-momentum, or $x$- and $z$-momentum combined, or both $x$- and $z$-momentum as well as volume fraction $\alpha$, when forcing to reference values for the undisturbed, calm free surface.
The results point out the necessity to adjust the tunable parameters according to the wave parameters. 
Finally, Appendixes A, B  and C discuss common forcing zone implementations, the optimum blending-in of the forcing source terms and the convergence of the analytical solution for the discretized problem towards the analytical solution of the continuous problem.

\section{Wave damping using forcing zones}
\label{SECforc}
This work considers fluid flow governed by the equation for mass conservation, the three equations for momentum conservation and the equation for the volume fraction, which describes the distribution of the phases: 
\begin{equation}
\frac{\mathrm{d}}{\mathrm{d} t} \int_{V} \rho \ \mathrm{d}V + \int_{S} \rho (\textbf{v} - \textbf{v}_{g} ) \cdot \textbf{n} \ \mathrm{d}S =  0 \quad ,
\label{EQconti}
\end{equation}
\begin{align}
\frac{\mathrm{d}}{\mathrm{d} t} \int_{V} \rho u_{i} \ \mathrm{d}V 
+ \int_{S} \rho u_{i} (\textbf{v} - \textbf{v}_{g} ) \cdot \textbf{n} \ \mathrm{d}S =  \nonumber \\ 
\int_{S} (\tau_{ij}\textbf{i}_{j} - p\textbf{i}_{i}) \cdot \textbf{n} \ \mathrm{d}S 
+ \int_{V} \rho \textbf{gi}_{i} \ \mathrm{d}V + \int_{V} \rho q_{i} \ \mathrm{d}V \quad ,
\label{EQnavier_stokes}
\end{align}
\begin{equation}
\frac{\rm d}{{\rm d} t} \int_{V} \alpha \ \mathrm{d}V + \int_{S} \alpha (\textbf{v} - \textbf{v}_{\rm g} ) \cdot \textbf{n} \ \mathrm{d}S =  \int_V q_{\alpha} \ \mathrm{d}V \quad .
\label{EQtransport_alpha}
\end{equation}
Here $V $ is the control volume (CV) bounded by the closed surface $\mathrm{S}$, \textbf{v} is the velocity vector of the fluid with the Cartesian components $u_{i}$, $\textbf{v}_{g} $ is the grid velocity, \textbf{n} is the unit vector normal to $S$ and pointing outwards, $t$ is time, $p$ is the pressure, $\rho$ are fluid density, $\tau_{ij}$ are the components of the viscous stress tensor,  \textbf{i}$_{j}$ is the unit vector in direction $ x_{j} $, with volume fraction $ \alpha $ of water.
Unless severe wave breaking occurs,  the propagation of ocean waves is an approximately inviscid phenomenon. Thus the results in this work apply regardless which formulation for $\tau_{ij}$ is chosen or whether it is neglected altogether.

Undesired wave reflections can be minimized by applying   source terms for volume fraction, $q_{\alpha}$, and momentum, $ q_{i} $, as
\begin{equation}
q_{\alpha} =\gamma   b(x) \left( \alpha_{\mathrm{ref}} -\alpha \right) \quad ,
\label{EQmassdamp}
\end{equation}
\begin{equation}
q_{\rm i} = \gamma  b(x) (u_{i,\mathrm{ref}}-u_{i})\quad ,
\label{EQmomdamp}
\end{equation}
with reference volume fraction $ \alpha_{\mathrm{ref}} $,  reference velocity component $u_{i,\mathrm{ref}} $,  forcing strength $\gamma$ and blending function $b(x)$. 

The unit of  $\gamma$ is  $\left[ \mathrm{rad/s} \right]$. It regulates the magnitude with which the solution at a given cell is forced against the reference solution. The optimum value for $\gamma$ is case-dependent. 
Apart from Eq. (\ref{EQmomdamp}), there exist also source terms which are not directly proportional to the forced quantity; these are discussed in Appendix A.

If the reference solution is the hydrostatic solution for the undisturbed free surface (e.g. $u_{i,\mathrm{ref}}=0$), then the forcing can be interpreted as a '\textit{wave damping}'.

\begin{figure}[H]
\begin{center}
\includegraphics[width=0.8\linewidth]{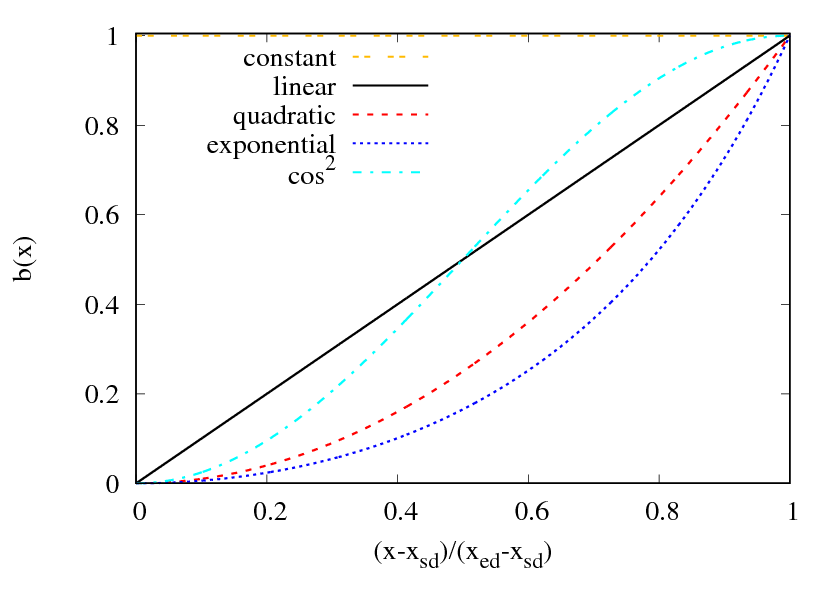}
\end{center}
\caption{Different blending functions $b(x)$ over location in forcing zone } \label{FIGblend}
\end{figure}

The blending term $b(x)$ regulates the distribution of the source term over the domain, where $x$ is the wave propagation direction.  Many different types of blending functions can be applied. 
Figure \ref{FIGblend} shows common blending functions, such as constant blending 
\begin{equation}
b(x) = 1 \quad ,
\label{EQblendconst}
\end{equation}
linear blending 
\begin{equation}
b(x) = \frac{x - x_{\rm sd}}{x_{\rm ed} - x_{\rm sd}} \quad ,
\label{EQblendlin}
\end{equation}
quadratic blending
\begin{equation}
b(x) = \left(\frac{x - x_{\rm sd}}{x_{\rm ed} - x_{\rm sd}}\right)^{2} \quad ,
\label{EQblendquad}
\end{equation}
 cosine-square blending
\begin{equation}
 b(x) = \cos^{2}\left( \frac{\pi}{2} +\frac{\pi}{2}\frac{x - x_{\rm sd}}{x_{\rm ed} - x_{\rm sd}} \right) \quad ,
\label{EQblendcos2}
\end{equation}
with start coordinate $ x_{\rm sd} $  and end coordinate $ x_{\rm ed} $ of the forcing zone. The thickness of the forcing zone is $x_{\mathrm{d}}=| x_{\rm ed} -x_{\rm sd}|$. Though at present it is unknown which blending function would be optimal, generally higher order blending functions are preferred, since they proved more effective in several investigations (e.g. Israeli and Orszag (1981)). Unless mentioned otherwise, an exponential blending is used in this work:
\begin{equation}
b(x) = \left( \frac{e^{\left( \frac{x - x_{\rm sd}}{x_{\rm ed} - x_{\rm sd}} \right)^{2}} - 1}{e^{1} - 1} \right) \quad .
\label{EQblendexp}
\end{equation}

Peri\'c and Abdel-Maksoud (2016) showed that forcing strength $\gamma$ and forcing zone thickness $x_{\mathrm{d}}$ scale as

\fbox {
    \parbox{\linewidth}{
\begin{equation}
\gamma \propto \omega, \quad x_{\mathrm{d}} \propto \lambda \quad , 
\label{EQscale}
\end{equation}
    }
}
with angular wave frequency $\omega=\frac{2\pi}{T}$ and wavelength $\lambda$. Thus to achieve the same reflection coefficient and similar free-surface elevations within the forcing zones when running two flow simulations, the parameters $\gamma$ and $x_{\mathrm{d}}$  for the second simulation have to be adjusted as
\begin{equation}
x_{{\rm d}} = x_{{\rm d, ref}} \cdot \frac{\lambda}{\lambda_{{\rm ref}}} \quad ,
\label{EQxdscale}
\end{equation}
\begin{equation}
\gamma = \gamma_{{\rm ref}} \cdot \frac{\omega}{\omega_{{\rm ref}}} \quad ,
\label{EQgammascale}
\end{equation}
where the $ x_{{\rm d, ref}}$, $ \lambda_{{\rm ref}} $, $\omega_{{\rm ref}}$ and $\gamma_{{\rm ref}} $ are the corresponding parameters from the first simulation.

Peri\'c and Abdel-Maksoud (2016) demonstrated that the optimum forcing strength $\gamma_{\mathrm{opt}}$  and the corresponding reflection coefficient $C_{\mathrm{R}}$ can be determined with comparatively low computational effort for a given wave period $T$, by running 2D flow simulations for different forcing strength $\gamma$. Via Eqs. (\ref{EQxdscale}) and (\ref{EQgammascale}) $\gamma_{\mathrm{opt}}$ can then be determined for any wave, for which the same blending $b(x)$ and zone thickness $x_{\mathrm{d}}/\lambda$ relative to the wavelength are used. In this case, $\gamma_{\mathrm{opt}}$ can for practical purposes be considered independent of the wave steepness, independent of the discretization (time step, order and choice of discretization scheme, mesh size), and roughly independent of  $x_{\mathrm{d}}$  in the interval $1\lambda \leq x_{\mathrm{d}} \leq 2 \lambda$; for example for blending according to Eq. (\ref{EQblendexp}), Peri\'c and Abdel-Maksoud (2016) obtained  $\gamma_{\mathrm{opt}} \approx \pi \omega$, which gives  $C_{\mathrm{R}} < 1\%$ for $ x_{\mathrm{d}} = 2\lambda$. 
A drawback of this approach is that, for a different blending $b(x)$ or zone thickness (i.e. $  x_{\mathrm{d}} <1\lambda$ or $  x_{\mathrm{d}} > 2\lambda$), the above calibration process has to be carried out again.

\section{Theory for predicting wave reflection coefficients}
\label{SECtheory}
\begin{figure}[H]
\begin{center}
\includegraphics[width=\linewidth]{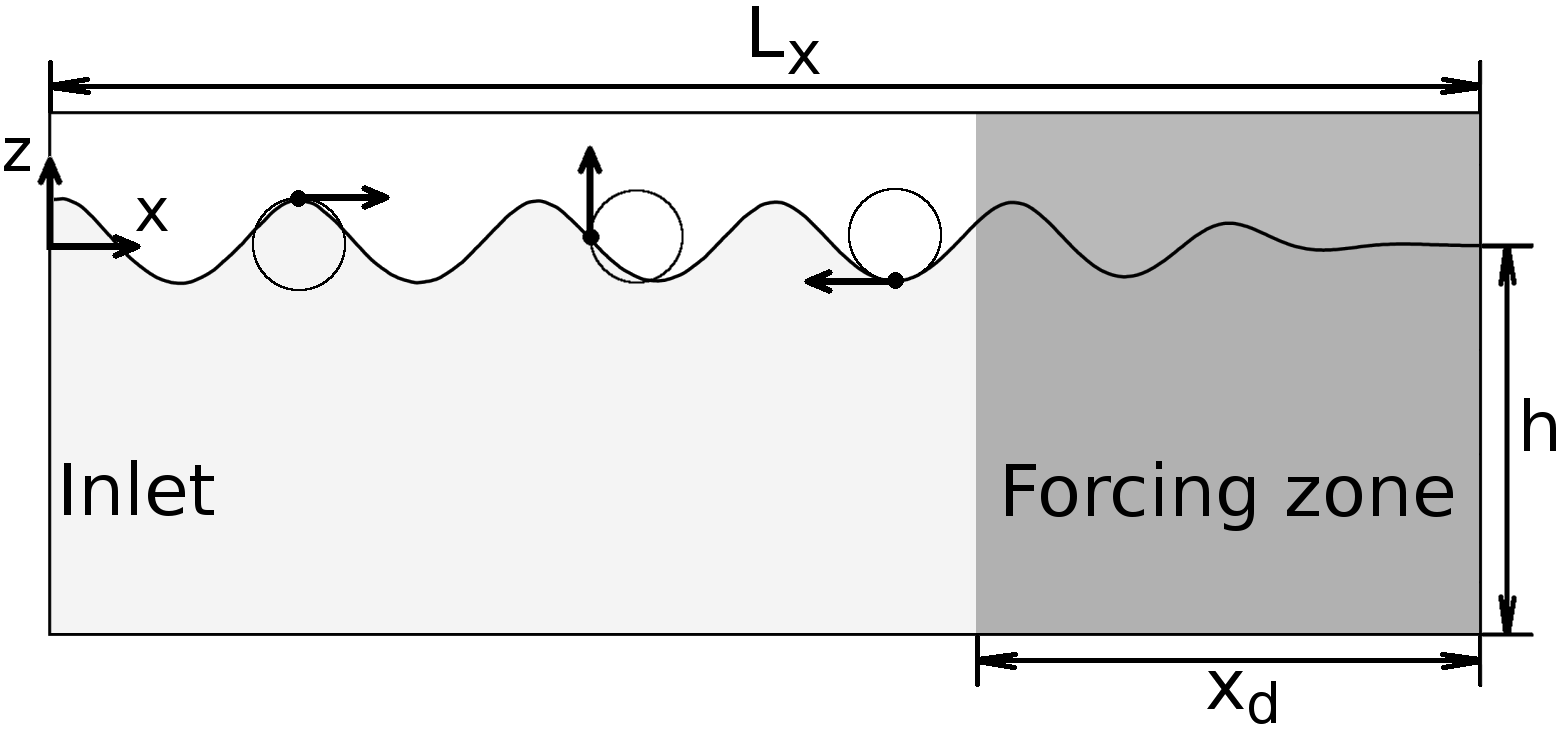}
\end{center}
\caption{Solution domain filled with air (white) and water (light gray, water depth $h$), velocity inlet at $x=0$ and forcing zone (shaded dark gray) with thickness $x_{\mathrm{d}}$; three fluid particles (black dots) are sketched with their particle paths (circles) and velocity vectors (arrows)} \label{FIGdom}
\end{figure}

This section considers the propagation of long-crested free-surface waves. As illustrated in Fig. \ref{FIGdom}, the waves are generated at $x=0$, and travel towards boundary $x=L_{x}$, to which a forcing zone with thickness $x_{\mathrm{d}}$ is attached. The coordinate system origin lies at the level  of the undisturbed free surface on the wave generating boundary  as illustrated in Fig. \ref{FIGdom}. The $x$-direction points in the wave  propagation direction and the $z$-direction is normal to the undisturbed free surface pointing away from the liquid phase. 

Outside the forcing zone, the waves fulfill the one-dimensional wave equation
\begin{equation}
\psi_{tt}=  c^{2} \psi_{xx}\quad .
\label{EQwave}
\end{equation}
with location $x$ in wave propagation direction, velocity stream function $\psi$, and phase velocity $c$. 

According to linear wave theory in the complex plane, the velocity stream function for an undamped wave is
\begin{equation}
 \psi =  \frac{H \omega}{2k} \frac{\sinh(k(z+h))}{\sinh(kh)}  \mathrm{e}^{i(-\omega t + k x)}\quad ,
\label{EQvelstreamBasic}
\end{equation}
with wave height $H$, angular wave frequency $\omega=2\pi/T$, wave period $T$, wave number $k=2\pi/\lambda$, wavelength $\lambda$, and water depth $h$.
Horizontal and vertical velocity components $u$ and $w$ are
\[u = \psi_{z}\quad ,\]
\[w = -\psi_{x}\quad .\]
Horizontal and vertical particle  displacements $\mathcal{X}$ and $\mathcal{Z}$ are
\begin{equation}
\mathcal{X} =   i\frac{H}{2} \frac{\cosh(k(z+h))}{\sinh(kh)}\mathrm{e}^{i(-\omega t + k x)}\quad ,
\label{EQpartpathx}
\end{equation}
\begin{equation}
\mathcal{Z} =   \frac{H}{2} \frac{\sinh(k(z+h))}{\sinh(kh)}\mathrm{e}^{i(-\omega t + k x)}\quad .
\label{EQpartpathz}
\end{equation}

Thus outside the forcing zone, Eq. (\ref{EQwave}) is equivalent to 
\begin{equation}
\mathcal{X}_{tt}=  c^{2} \mathcal{X}_{xx}\quad .
\label{EQwavexsi}
\end{equation}
A detailed discussion of linear wave theory can be found in Clauss et al. (1992).

\subsection{Forcing of $x$-momentum}
\label{SECtheoryxmom} 
Waves can be damped by attaching a forcing zone to the domain, in which the horizontal particle velocity $u=\mathcal{X}_{t}$ is forced to a reference velocity   $u=\mathcal{X}_{t,\mathrm{ref}}$
\begin{equation}
 \mathcal{X}_{tt}=  c^{2} \mathcal{X}_{xx} +  \underbrace{\gamma  b(x) (\mathcal{X}_{t,\mathrm{ref}} - \mathcal{X}_{t})}_{q}  \quad ,
\label{EQwavewithdampingX}
\end{equation}
with forcing strength $\gamma$ and blending function $b(x)$.
The last term in Eq. (\ref{EQwavewithdampingX}), $q$, is equivalent to the forcing source term in Eq. (\ref{EQmomdamp}). Since in this work  $\mathcal{X}_{t,\mathrm{ref}} = 0$,  Eq. (\ref{EQwavewithdampingX}) corresponds to a 'classical' absorbing layer formulation.
Further, Eq. (\ref{EQwavewithdampingX}) is equivalent to 
\begin{equation}
 \psi_{tt}=  c^{2} \psi_{xx} + \gamma  b(x) (\psi_{t,\mathrm{ref}} - \psi_{t})  \quad ,
\label{EQwavewithdampingpsi}
\end{equation}
with temporal derivative of the reference stream function $\psi_{t,\mathrm{ref}} = 0$.

Inserting Eq. (\ref{EQvelstreamBasic})  into  Eq. (\ref{EQwave}) gives the wave number \textit{outside the forcing zone}
\begin{equation}
k= \sqrt{\frac{\omega^{2} }{c^{2}} }  = \frac{\omega}{c} \quad .
\label{EQk}
\end{equation}

Inserting Eq. (\ref{EQvelstreamBasic}) for piecewise-constant blending $b(x)$ into  Eq. (\ref{EQwavewithdampingpsi}) gives the wave number \textit{inside the forcing zone}
\begin{equation}
 k= \sqrt{\frac{\omega^{2}}{c^{2}} + i \frac{  \omega \gamma  b(x) }{c^{2}} } \quad .
\label{EQkstar}
\end{equation}
Thus  inside the forcing zone, the wave number contains an additional imaginary part which damps the wave amplitude but does not change the wavelength.

\begin{figure}[H]
\begin{center}
\includegraphics[width=\linewidth]{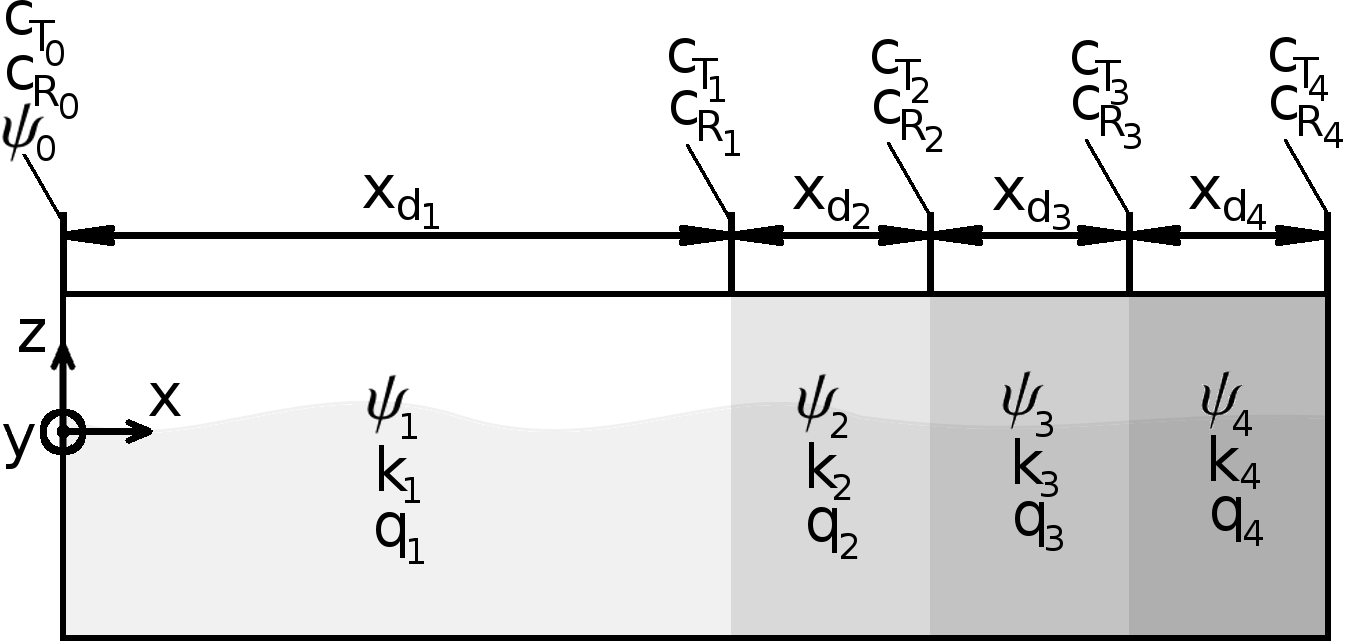}
\end{center}
\caption{Example of a solution domain decomposed into $4$ zones; within each zone $j$ of thickness $x_{\mathrm{d},j}$ holds stream function $\psi_{j}$, wave number $k_{j}$ and damping term $q_{j}$; at each interface $j$ between two zones,  transmission and reflection coefficients are $C_{\mathrm{T}_{j}}$ and $C_{\mathrm{R}_{j}}$} \label{FIGdomwith4xdzones}
\end{figure}

The analytical solution to the problem of determining the reflection coefficient for a wave entering a forcing zone according to Eqs. (\ref{EQwave}) and (\ref{EQwavewithdampingpsi}) is obtained by an approach somewhat analogous to the way in which forcing source terms are applied in numerical flow simulations on finite grids. 
The solution domain is discretized into a finite number of cells as illustrated in Fig. \ref{FIGdomwith4xdzones}. Each cell $j$ corresponds to a forcing zone, within which the stream function is given by $\psi_{j}$, and the complex wave number $k_{j}$ has a constant value. For this, $b(x)$ is evaluated at the cell center:
\begin{equation}
 k_{j}= \sqrt{\frac{\omega^{2}  + i \omega \gamma  b( \sum_{n=1}^{j-1} x_{\mathrm{d}_{n}} + \frac{1}{2}x_{\mathrm{d}_{j}}) }{c^{2}} } \quad ,
\label{EQkstar}
\end{equation}
with thickness $x_{\mathrm{d},j}$ of  zone $j$; $x_{\mathrm{d},j}$ is equivalent to the size of the cell in $x$-direction. Thus the damping is constant within every zone.
Reflection and transmission may occur at every interface between two cells. 

The benefit of this approach is that even non-continuous blending functions and the influence of the discretization  can be considered.
With increasing resolution, the theoretical results are expected to converge to the solution of the continuous problem.  
The latter is not derived here, since for practical purposes only the analytical solution to the discretized problem is of interest. 
In this manner, the problem remains linear and the solution can be derived as follows.

Consider a wave propagating in positive $x$-direction. The wave is generated at $x=0$ following the coordinate system in Fig. \ref{FIGdomwith4xdzones}. Let the stream function at $x=0$ be 
\begin{align}
\psi_{0} = \frac{H\omega}{2k_{0}} \frac{\sinh (k_{0}(z+h))}{\sinh (k_{0}h)}  \mathrm{e}^{i(-\omega t)} \quad ,
\label{EQpsi0}
\end{align} 
with wave height $H$, angular wave frequency $\omega$, wave number $k_{0}$, vertical coordinate $z$, and time $t$. Set the transmission coefficient $C_{\mathrm{T}_{0}} = 1$ and the  reflection coefficient $C_{\mathrm{R}_{0}} = 0$, thus the 'inlet' boundary is perfectly transparent, and waves propagating through it in negative $x$-direction will be fully transmitted without reflection. 

For illustration, a domain with $4$ zones is  depicted in  Fig. \ref{FIGdomwith4xdzones}. Let the wave number  $k_{1}=2\pi/\lambda_{1}$ within zone $1$ equal the wave number $k_{0}=2\pi/\lambda_{0}$ of the  wave generated at $x=0$, where  $\lambda_{0}$ and  $\lambda_{1}$ are the corresponding wavelengths. Thus within the first zone  $0 \leq x \leq x_{\mathrm{d}_{1}}$, there is no wave damping, i.e. $q(x) = 0$. At the end of the domain, i.e. at $x =\sum_{n=1}^{4}  x_{\mathrm{d}_{n}} $, the boundary is perfectly reflecting (a typical 'wall boundary condition' in computational fluid dynamics), so the transmission coefficient $C_{\mathrm{T}_{4}} = 0$ and the  reflection coefficient $C_{\mathrm{R}_{4}} = 1$.
 Within each zone in zones $2$ to $4$, the damping is constant, i.e. $q(x)=q_{j}=-\gamma  b( \sum_{n=1}^{j-1} x_{\mathrm{d}_{n}} + \frac{1}{2}x_{\mathrm{d}_{j}}) \mathcal{X}_{t} $, yet $q_{2}$ to $q_{4}$ may be of different magnitude, which is indicated through the different shading of the zones in  Fig. \ref{FIGdomwith4xdzones}.

By requiring that the particle displacements and velocities must be continuous at every interface between two zones, as they should be at the interfaces between two cells in a flow simulation, the periodic solution is obtained.

For a domain with $j$ zones, the general solution for the velocity stream function $\psi(x)$ within zone $j>0$ can be written as a sum of a right-going (incoming) and a left-going (reflected) wave component
\begin{align}
 \psi_{j} =  \psi_{0}  \left( \prod_{n=0}^{j-1} C_{\mathrm{T}_{n}} \right)
\cdot  \biggl[ \mathrm{e}^{i \left( \sum_{n=1}^{j-1} k_{n} x_{\mathrm{d}_{n}} + k_{j} \left( x -  \sum_{n=1}^{j-1} x_{\mathrm{d}_{n}  }   \right) \right)} \nonumber \\   - C_{\mathrm{R}_{j}} \mathrm{e}^{i \left(  \sum_{n=1}^{j-1} k_{n} x_{\mathrm{d}_{n}}  + k_{j}2x_{\mathrm{d}_{j}} - k_{j} \left( x -  \sum_{n=1}^{j-1} x_{\mathrm{d}_{n} }   \right) \right)} 
\biggr]  \quad ,
\label{EQXjgeneralt2}
\end{align} 
with $\psi_{0}$ according to Eq. (\ref{EQpsi0}).
The derivative of $ \psi_{j}$ with respect to wave propagation direction $x$ is thus
\begin{align}
 \psi_{x,j} =  i k_{j} \psi_{0} \left( \prod_{n=0}^{j-1} C_{\mathrm{T}_{n}} \right) 
\cdot  \biggl[ \mathrm{e}^{i \left( \sum_{n=1}^{j-1} k_{n} x_{\mathrm{d}_{n}} + k_{j} \left( x -  \sum_{n=1}^{j-1} x_{\mathrm{d}_{n}  }   \right) \right)} \nonumber \\  + C_{\mathrm{R}_{j}} \mathrm{e}^{i \left(  \sum_{n=1}^{j-1} k_{n} x_{\mathrm{d}_{n}}  + k_{j}2x_{\mathrm{d}_{j}} - k_{j} \left( x -  \sum_{n=1}^{j-1} x_{\mathrm{d}_{n} }   \right) \right)} 
\biggr]  \quad .
\label{EQXjgeneralt2x}
\end{align} 

\textbf{Requirement 1:} At the interface between zones $j$ and $j+1$, the solution for  the zone on the left, i.e.  $\psi_{j}$, must equal  the  solution for the zone on the right, i.e.  $\psi_{j+1}$: 
\begin{align}
 \left[ \psi_{j} =  \psi_{j+1} \right]_{x=\sum_{n=1}^{j} x_{\mathrm{d}_{n}}} \quad .
 \label{EQBC1t2}
\end{align} 

Inserting Eq. (\ref{EQXjgeneralt2}) into Eq. (\ref{EQBC1t2}), dividing by 
\[   \psi_{0}  \left( \prod_{n=0}^{j-1} C_{\mathrm{T}_{n}} \right)\biggl[ \mathrm{e}^{i \left( \sum_{n=1}^{j-1} k_{n} x_{\mathrm{d}_{n}} \right) }  \biggr] \quad ,\]
 inserting 
\[ x - \sum_{n=1}^{j-1} x_{\mathrm{d}_{n}} = x_{\mathrm{d}_{j} }  \quad , \quad x - \sum_{n=1}^{j} x_{\mathrm{d}_{n}} = 0 \quad , \]
and rearranging for $ C_{\mathrm{T}_{j}}$ gives
\begin{framed}
\begin{align}
 C_{\mathrm{T}_{j}} =  \frac{ 1  - C_{\mathrm{R}_{j}} }{  1 - C_{\mathrm{R}_{j+1} }  \mathrm{e}^{i \left(  k_{j+1}2x_{\mathrm{d}_{j+1} }  \right) } } 
 \quad .
\label{EQCtt2}
\end{align} 
\end{framed}

\textbf{Requirement 2:} At the interface between zones $j$ and $j+1$, the  spatial derivative with respect to $x$ of the solution for  the zone on the left, i.e.  $\partial \psi_{j}/\partial x = \psi_{x,j}$, must equal the spatial derivative with respect to $x$ of the solution for the zone on the right, i.e.  $\partial \psi_{j+1}/\partial x = \psi_{x,j+1}$: 
\begin{align}
 \left[ \psi_{x,j} =  \psi_{x,j+1} \right]_{x=\sum_{n=1}^{j} x_{\mathrm{d}_{n}}} \quad .
 \label{EQBC1t2x}
\end{align} 

Inserting Eq. (\ref{EQXjgeneralt2x}) into Eq. (\ref{EQBC1t2x}), dividing by 
\[  i \psi_{0}  \left( \prod_{n=0}^{j-1} C_{\mathrm{T}_{n}} \right)\biggl[ \mathrm{e}^{i \left( \sum_{n=1}^{j-1} k_{n} x_{\mathrm{d}_{n}} \right) }  \biggr] \quad ,\]
 inserting $ C_{\mathrm{T}_{j}}  $, and setting 
 \begin{align}
 \beta_{j+1} =  \frac{1 + C_{\mathrm{R}_{j+1}} \mathrm{e}^{i \left( k_{j+1}2 x_{\mathrm{d}_{j+1}} \right)}  }{ 1 - C_{\mathrm{R}_{j+1}}  \mathrm{e}^{i \left(  k_{j+1}2x_{\mathrm{d}_{j+1}} \right) }}\quad .
\label{EQCRjz2}
\end{align} 
gives
\begin{framed}
\begin{align}
  C_{\mathrm{R}_{j}} = \frac{  k_{j+1}\beta_{j+1} -  k_{j}   }{ k_{j+1}\beta_{j+1} + k_{j} }
 \quad ,
\label{EQCRjz2}
\end{align} 
\end{framed}

For practical purposes, mainly the 'global' reflection coefficient $  C_{\mathrm{R}} $ is of interest, which is the ratio of the amplitude of the wave, which is reflected back into the solution domain, to the amplitude of the wave, which enters the forcing zone.
  The global reflection coefficient $  C_{\mathrm{R}} $ corresponds to the magnitude of  $  C_{\mathrm{R}_{j}} $ at the interface to the forcing zone,  which depends on all $  C_{\mathrm{R}_{j}} $ inside the whole forcing zone. So if the forcing zone starts at zone $1$, then
\begin{framed}
 \begin{align}
   C_{\mathrm{R}} = | C_{\mathrm{R}_{1}} | = \sqrt{\mathrm{Re}\{C_{\mathrm{R}_{1}}\}^{2} + \mathrm{Im}\{C_{\mathrm{R}_{1}}\}^{2}}
 \quad ,
\label{EQCRglobal}
\end{align} 
\end{framed}
where $ \mathrm{Re}\{ X\} $ and $ \mathrm{Im}\{ X\} $ denote the real and the imaginary part of the complex number $X$.

 Since for the complex exponential  functions in this case the order of derivatives and integrals can be exchanged, when setting integration constants to zero one obtains
\[  u_{j} = \psi_{z,j} = k_{0}  \frac{\cosh (k_{0}(z+h))}{\sinh (k_{0}(z+h))}\psi_{j}\quad , \]
\[  \mathcal{X}_{j} = \int \psi_{z,j}\, \mathrm{d}t =  \frac{k_{0}}{-i\omega}  \frac{\cosh (k_{0}(z+h))}{\sinh (k_{0}(z+h))}\psi_{j}\quad , \]
\[  w_{j} = -\psi_{x,j} \quad , \]
\[ \mathcal{Z}_{j} = \int -\psi_{x,j}\, \mathrm{d}t =  \frac{1}{i\omega} \psi_{x,j}\quad . \]
Thus $ \psi_{j}$, $ \psi_{x,j}$ and and all particle displacements and velocities are continuous throughout the domain.

\subsection{Forcing of volume fraction $\alpha$, $x$- and $z$-momentum}
\label{SECtheoryxza}
Combinations of forcing of volume fraction $\alpha$, $x$- and $z$-momentum can be described by the above theory, when $\gamma$ is adjusted as outlined in the following. 

Assume that the equations for volume fraction, $x$- and $z$-momentum are coupled, so that forcing of one equation acts on the other equations immediately.

According to linear wave theory, the  average energy $\bar{E} = \int_{0}^{\lambda} \int_{-h}^{\eta} \int_{0}^{1\, \mathrm{m}} E \, \mathrm{d}y\mathrm{d}z\mathrm{d}x$ in a regular deep-water free-surface wave can be subdivided as
\begin{equation}
\bar{E} = \bar{E}_{\mathrm{pot}} + \underbrace{\bar{E}_{\mathrm{kin}}}_{=\bar{E}_{\mathrm{kin,x}} + \bar{E}_{\mathrm{kin,z}}}  \quad ,
\label{EQEsubdivision1}
\end{equation}
where
\begin{equation}
\bar{E}_{\mathrm{pot}} = \bar{E}_{\mathrm{kin}} \quad \mathrm{and }\quad \bar{E}_{\mathrm{kin,x}} = \bar{E}_{\mathrm{kin,z}} = \frac{1}{2} \bar{E}_{\mathrm{pot}}\quad ,
\label{EQEsubdivision2}
\end{equation}
with location of the free surface $\eta$ above still water level, water depth $h$, average potential and kinetic energy $\bar{E}_{\mathrm{pot}}$ and $\bar{E}_{\mathrm{kin}}$, and the $x$- and $z$-component $ \bar{E}_{\mathrm{kin,x}}$ and $ \bar{E}_{\mathrm{kin,z}} $ of the average kinetic energy.

Average potential and kinetic energy in the wave have the same magnitude. Therefore applying forcing with same $\gamma$   to both potential and kinetic energy ($q_{\alpha}$, $q_{\mathrm{x}}$ and $q_{\mathrm{z}}$) produces a forcing of twice the strength as when applying forcing with same $\gamma$ only to the kinetic energy (i.e. $q_{\mathrm{x}}$ and $q_{\mathrm{z}}$); thus $\gamma_{(E_{\mathrm{pot}}+E_{\mathrm{kin}})}= \frac{1}{2}\gamma_{(E_{\mathrm{kin}})}$. 

Similarly, since the $x$- and $z$-components of the kinetic energy have on average the same magnitude (i.e. $\bar{E}_{\mathrm{kin,x}} = \bar{E}_{\mathrm{kin,z}} $), applying forcing to only one of these (i.e. either $q_{\mathrm{x}}$ or $q_{\mathrm{z}}$) shifts the optimum of $\gamma$ even further to the right. Therefore, it generally holds that\\
\fbox {
    \parbox{\linewidth}{
\begin{equation}
\gamma_{(E_{\mathrm{kin,z}})}=\gamma_{(E_{\mathrm{kin,x}})} = 2\gamma_{(E_{\mathrm{kin}})}= 2\gamma_{(E_{\mathrm{pot}})}= 4\gamma_{(E_{\mathrm{pot}}+E_{\mathrm{kin}})} \quad ,
\label{EQcomponents}
\end{equation} 
    }
}
with forcing strengths for the cases of forcing of $w$-velocity, $ \gamma_{(E_{\mathrm{kin,z}})}$, forcing of $u$-velocity, $\gamma_{(E_{\mathrm{kin,x}})}$, forcing of both $u$- and $w$-velocities,  $\gamma_{(E_{\mathrm{kin}})}$, as well as forcing of $u$- and $w$-velocities and volume fraction $\alpha$, $\gamma_{(E_{\mathrm{pot}}+E_{\mathrm{kin}})}$.

Therefore, the theory from Sect. \ref{SECtheory}  can be used for the other forcing source terms as well, given that $\gamma$ is adjusted according to Eq. (\ref{EQcomponents}); thus when simultaneously forcing $\alpha$, $x$- and $z$-momentum, each with forcing strength $\gamma = 20\, \mathrm{rad/s}$, then the theoretical prediction for the reflection coefficient $C_{\mathrm{R}}$ is the same as when using the theory from Sect. \ref{SECtheoryxmom} with $\gamma=80\, \mathrm{rad/s}$.

In the following sections, the theoretically predicted reflection coefficients will be compared with those obtained from flow simulations based on the solution of the Navier-Stokes equations.

\section{Simulation setup}
\label{SECsetup}
In the flow simulations in this work, a regular long-crested free-surface wave train is created and propagates in positive $x$-direction towards a forcing zone as sketched in Fig. \ref{FIGdom}, where it is partly reflected and partly absorbed. The wave has height $H=0.16\, \mathrm{m}$, period $T=1.6\, \mathrm{s}$, wavelength $\lambda\approx 4\, \mathrm{m}$ and is moderately non-linear (steepness $H/\lambda $ is $\approx 29\%$ of maximum steepness). Deep water conditions apply ($h/\lambda> 0.5$).

The commercial flow solver STAR-CCM+ version 11.06.010-R8 from Siemens (formerly CD-adapco) is used for the simulations. 
The governing Eqs. are (\ref{EQconti}) to (\ref{EQtransport_alpha}). The volume of fluid (VOF) method is used to account for the two fluid phases, liquid water and gaseous air, using the High Resolution Interface Capturing scheme (HRIC) as given in Muzaferija and Peri\'c (1999).
The governing equations are applied to each cell and discretized according to the finite volume method. All integrals are approximated by the midpoint rule. The interpolation of variables from cell center to face center and the numerical differentiation are performed using linear shape functions, leading to approximations of second order. The integration in time is based on assumed quadratic variation of variables in time, which is also a second-order approximation. Each algebraic equation contains the unknown value from the cell center and the centers of all neighboring cells with which it shares common faces. The resulting coupled equation system is then linearized and solved by the iterative STAR-CCM+ implicit unsteady segregated solver, using an algebraic multigrid method with Gauss-Seidel relaxation scheme, V-cycles for pressure and volume fraction of water, and flexible cycles for velocity calculation. The under-relaxation factor is $0.9$ for velocities and volume fraction and $0.4$ for pressure.
For each time step, eight iterations are performed; one iteration consists of solving the governing equations for the velocity components, the pressure-correction equation (using the SIMPLE method for collocated grids to obtain the pressure values and to correct the velocities) and the transport equation for the volume fraction of water.
Further information on the discretization of and solvers for the governing equations can be found in  Ferziger and Peri\'c (2002) or the STAR-CCM+ software manual.

The forcing approaches from Sect. \ref{SECforc} are used to minimize wave reflections. Simulations are either run with forcing of $x$-momentum ($q_{\mathrm{x}}$), of $z$-momentum ($q_{\mathrm{z}}$), of both $x$- and $z$-momentum ($q_{\mathrm{x}}$, $q_{\mathrm{z}}$), or of volume fraction $\alpha$ and $x$- and $z$-momentum ($q_{\alpha}$, $q_{\mathrm{x}}$, $q_{\mathrm{z}}$). The forcing zone has exponential blending according to Eq. (\ref{EQblendexp}). Forcing zone thicknesses $x_{\mathrm{d}}$ in the range between $0.25\lambda$ and $2\lambda$ are investigated.

The simulations are run quasi-2D, i.e. with only one layer of cells in $y$-direction and symmetry boundary conditions applied to the $y$-normal boundaries. 
The domain has dimensions $0 \leq x \leq L_{x} $ and $-h \leq z \leq 0.5\lambda$, with domain length $L_{x}=6\lambda$ and water depth $h= 4.5\lambda$. 

The wave is generated by prescribing the volume fraction and velocities according to Fenton's (1985)  $5^{\mathrm{th}}$-order Stokes theory  at the inlet boundary ($x=0$). 
Due to the Stokes drift of the fluid particles in the wave, the inlet produces a net mass flux into the domain, which acts to raise the mean water level; by prescribing the hydrostatic pressure at either the  boundary to which the forcing zone is attached ($x=L_{x}$) or the bottom boundary ($z=-h$),  such an undesired accumulation of mass can be avoided. Peri\'c and Abdel-Maksoud (2015) found that the first option can lead to low-frequency fluctuations in the amount of mass within the domain, since the forcing zone delays how the pressure boundary regulates the amount of mass within the domain; prescribing the pressure at the domain bottom resolved this problem. In practice however, it is far more common to prescribe the pressure at the boundary to which the forcing zone is attached.
For this reason, all cases investigated in this work were simulated with pressure prescribed at boundary $x=L_{x}$. 

\begin{figure}[H]
\begin{center}
\includegraphics[width=\linewidth]{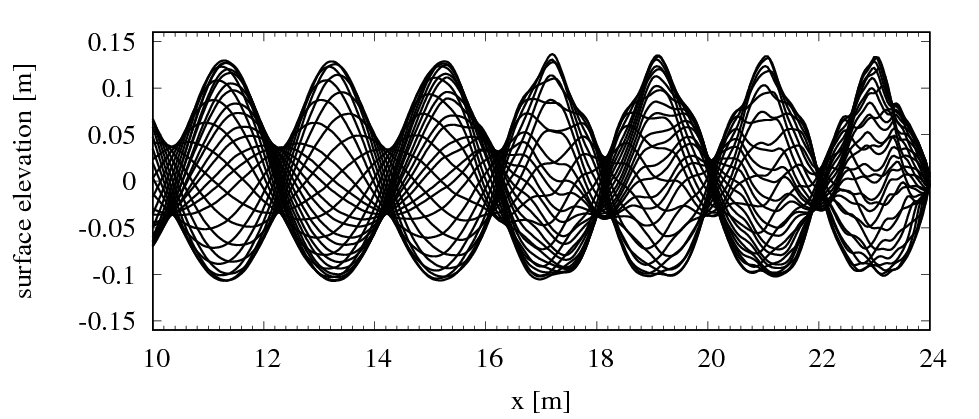}
\includegraphics[width=\linewidth]{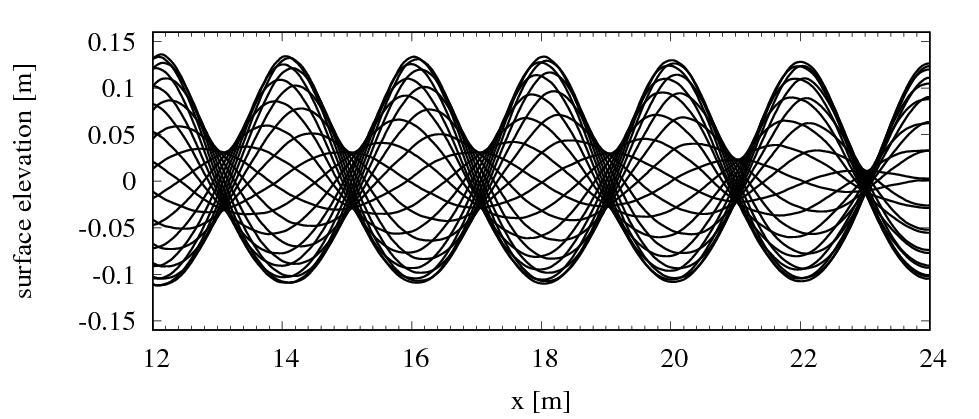}
\end{center}
\caption{Simulated free-surface elevation for  equally spaced time instances during one wave period, with vertical boundary $x=L_{x} = 24\, \mathrm{m}$ set as prescribed pressure  (top image, node at $x=24\, \mathrm{m}$) or wall boundary (bottom image, maximum amplification at $x=24\, \mathrm{m}$); results from Sect. \ref{SECresx} for $x_{\mathrm{d}}=2\lambda$ and weak damping ($\gamma=0.625\, \mathrm{rad/s}$)  } \label{FIGpoutposs}
\end{figure}

Additionally, the simulations for $q_{\mathrm{x}}$ and $q_{\mathrm{z}}$ with $x_{\mathrm{d}}=2\lambda$ were  rerun with pressure prescribed at boundary $z=-h$ and boundary  $x=L_{x}$ set to no-slip wall. The results confirmed that the choice for the pressure boundary does not significantly influence the reflection coefficients (on average $<1\%$ difference). 
Therefore in practice, the theory presented in this work can be used for both boundary choices. Yet from an academical point of view, there is a difference: as illustrated in Fig. \ref{FIGpoutposs}, prescribing the pressure at $x=L_{x}$ produces a node at that boundary; when the boundary condition is set to wall instead, there occurs a maximum amplification point. The theory in Sect. \ref{SECtheory} is derived for the latter case. Thus in Sect. \ref{SECresdisc}, comparisons of surface elevation with theory (Figs.  \ref{FIGxelevsim} and \ref{FIGzelevsim}) are given for the simulations with pressure prescribed at the domain bottom; all other results are given for pressure prescribed at $x=L_{x}$, because this approach is primarily used in practice and the observed difference in results between the two approaches were insignificant.

All remaining boundaries are no-slip walls. For further details on boundary conditions, see Ferziger and Peri\'c (2002).

\begin{figure}[H]
\begin{center}
\includegraphics[width=\linewidth]{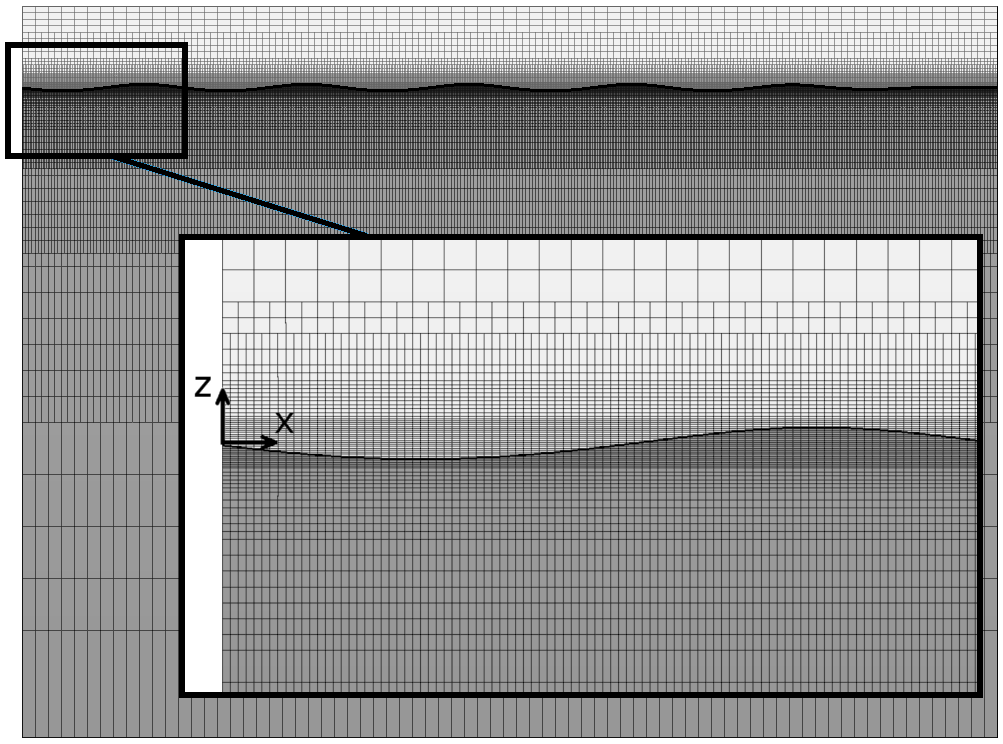}
\end{center}
\caption{Computational grid with close-up, showing the initialized location of the free surface (thick black curve) and of the liquid phase  (shaded dark gray)} \label{FIGgrid}
\end{figure}

The volume fraction and velocities in the solution domain are initialized according to Fenton's (1985) Stokes $5^{\mathrm{th}}$-order theory to reduce the simulation time. The domain is discretized using a rectilinear grid with local mesh refinement around the free surface.  The free surface stays at all times within the region of the finest mesh with $100$ cells per wavelength and $16$ cells per wave height. The computational grid, which consists of  $\approx 43\, 000$ cells, is shown in Fig. \ref{FIGgrid}. 

The total simulated time is $18\, \mathrm{s} \approx 11.3T$ with a time step of $\Delta t = T/1000$.

To calculate the reflection coefficient, the  free-surface elevation is written to a file at $80$ evenly spaced time intervals per wave period. From the elevation in an interval of $1.5\lambda$ adjacent to the forcing zone, the overall highest and lowest wave heights are obtained as $ H_{\mathrm{max}} $ and  $ H_{\mathrm{min}} $. From these, the reflection coefficient is calculated as in experiments after the procedure by Ursell et al. (1960) as
\begin{equation}
C_{\mathrm{R}} = \left(H_{\mathrm{max}} - H_{\mathrm{min}}\right)/\left(H_{\mathrm{max}} + H_{\mathrm{min}}\right) \quad . 
\label{EQcr}
\end{equation}
This simple approach contains only a small background noise ($\approx 1\%$, see Fig. \ref{SECresdisc}) due to the finite number of time intervals, which was considered accurate enough for the present purposes.

The reflection coefficient gives the ratio of the the wave heights of reflected to incoming wave; the ratio of the energy of the reflected ($E_{\mathrm{refl}}$) wave to the incoming ($E_{\mathrm{in}}$) wave is $E_{\mathrm{refl}}/E_{\mathrm{in}}\approx C_{\mathrm{R}}^{2}$, since according to linear wave theory the wave energy depends on the wave height squared. Thus a reflection coefficient of $C_{\mathrm{R}} = 10\%$ means that the forcing zone reflects $1\%$ of the  incoming wave energy.

\section{Results and discussion}
\label{SECresdisc}

In this section, reflection coefficients obtained in flow simulations using different forcing approaches and parameters are compared to theoretical predictions according to Sect. \ref{SECtheory}.

\subsection{Forcing of $x$-momentum}
\label{SECresx}
Flow simulations are carried out with the setup from Sect. \ref{SECsetup} and compared to the theory presented in Sect. \ref{SECtheory}. Forcing of $x$-momentum according to Eq. (\ref{EQmomdamp}) with exponential blending from Eq. (\ref{EQblendexp}) is used to damp waves with period $T=1.6\, \mathrm{s}$ and height $H=0.16\, \mathrm{m}$. Simulations are run for different forcing strengths ($0.625\, \mathrm{rad/s} \leq \gamma \leq 5120\, \mathrm{rad/s} $) and forcing zone thicknesses ($0.25\lambda \leq x_{\mathrm{d}} \leq 2\lambda$). 

\begin{figure}[H]
\includegraphics[width=\linewidth]{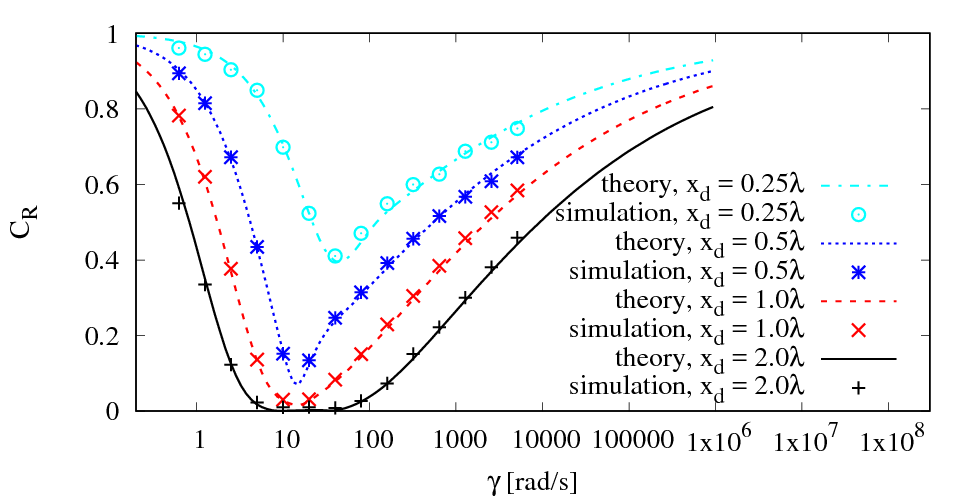} 
\caption{Reflection coefficient $C_{\mathrm{R}}$ over forcing strength $\gamma$ for forcing of $x$-momentum, given for different forcing zone thickness $x_{\mathrm{d}} $  according to simulation and theory } \label{FIGCrGammaXmom}
\end{figure}

Comparing simulation results and theory in Fig. \ref{FIGCrGammaXmom} shows that the theory can predict reflection coefficient $C_{\mathrm{R}}$ with high accuracy. 
Although the waves are moderately nonlinear ($\approx 29\%$ of maximum steepness), the deviations are small; this agrees with results from Peri\'c and Maksoud (2016), where deep-water waves up to $\approx 71\%$ of the maximum steepness were investigated, and the influence of wave steepness on the reflection coefficient in this range was found to be negligible for practical purposes ($<2\%$ difference), except for $\gamma$ roughly an order of magnitude below optimum, where the damping improved noticeably for the steeper wave.

\begin{figure}[H]
\includegraphics[width=\linewidth]{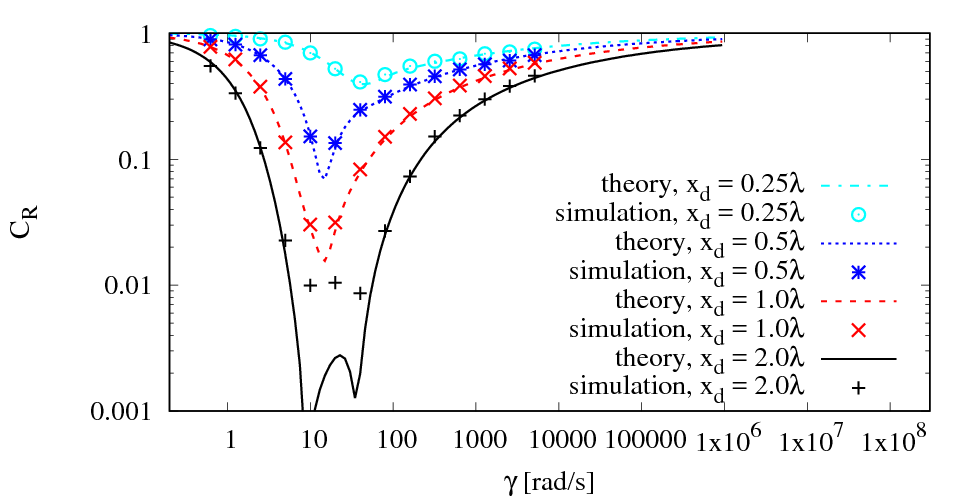} 
\caption{As Fig. \ref{FIGCrGammaXmom}, with log-scale for $C_{\mathrm{R}}$} \label{FIGCrGammaXmomLOG}
\end{figure}

The close-up on the optimum forcing regime in Fig. \ref{FIGCrGammaXmomLOG} shows that simulation results were not able to reproduce reflection coefficients below $\approx 1\%$, see e.g. the curve for $x_{\mathrm{d}}=2\lambda$ at $\gamma \approx 20\, \mathrm{rad/s}$ in Fig. \ref{FIGCrGammaXmomLOG}; this was attributed to a background noise in the approach for calculating $C_{\mathrm{R}}$, due to the finite number of time intervals used; possibly also the interpolation of the surface elevation on the finite grid limits the detection of $C_{\mathrm{R}}< 1\%$.

Further, while for thinner zones ($x_{\mathrm{d}}<1.5\lambda$) there is a single optimum for forcing strength $\gamma$, for thicker zones ($x_{\mathrm{d}}=2\lambda$) there can be more than one local optimum. This may have implications for the choice of $\gamma$ when damping irregular waves, which future research may show.

\begin{figure}[H]
\begin{center}
$\gamma=0.625\, \mathrm{rad/s}$\\
\includegraphics[width=\halffactor\linewidth]{images/xdamp/xelev0.625.png}\\
 $\gamma=2.5\, \mathrm{rad/s}$\\
\includegraphics[width=\halffactor\linewidth]{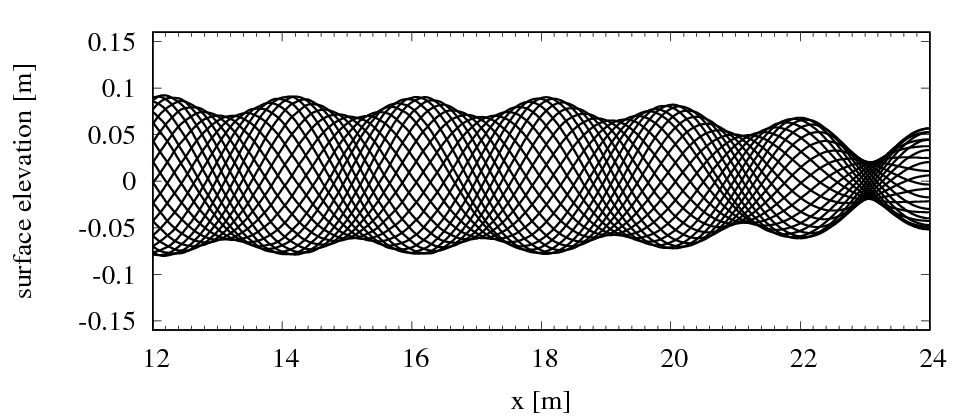}\\
 $\gamma=10\, \mathrm{rad/s}$\\
\includegraphics[width=\halffactor\linewidth]{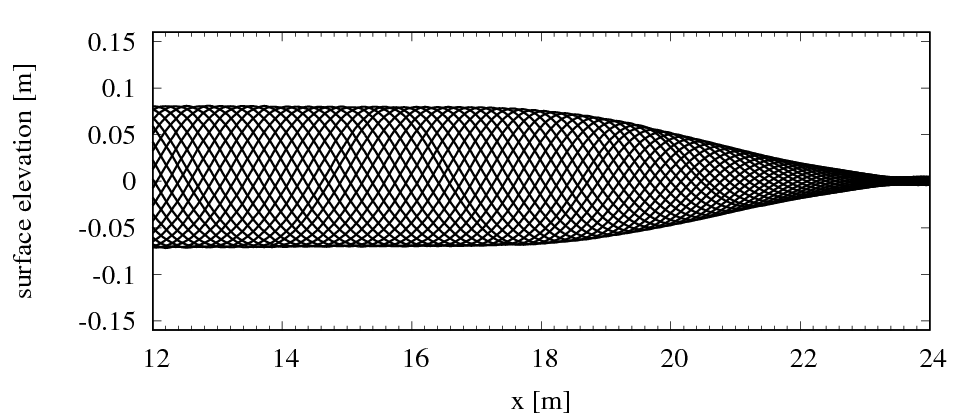}\\
 $\gamma=80\, \mathrm{rad/s}$\\
 \includegraphics[width=\halffactor\linewidth]{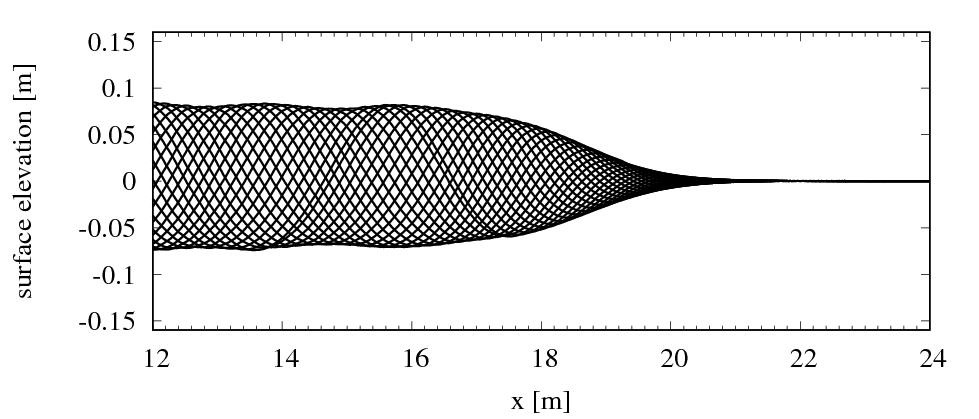}\\
 $\gamma=640\, \mathrm{rad/s}$\\
\includegraphics[width=\halffactor\linewidth]{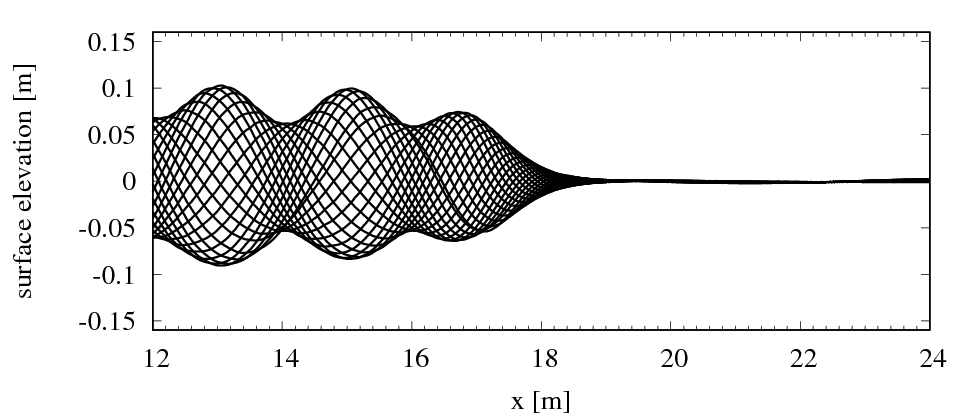}\\
\end{center}
\caption{Simulation results for surface elevation over $x$-coordinate in the vicinity of a forcing zone with thickness $x_{\mathrm{d}}=2\lambda$; evaluated for equally spaced time intervals during the last simulated period } \label{FIGxelevsim}
\end{figure}

\begin{figure}[H]
\begin{center}
$\gamma=0.625\, \mathrm{rad/s}$\\
\includegraphics[width=\halffactor\linewidth]{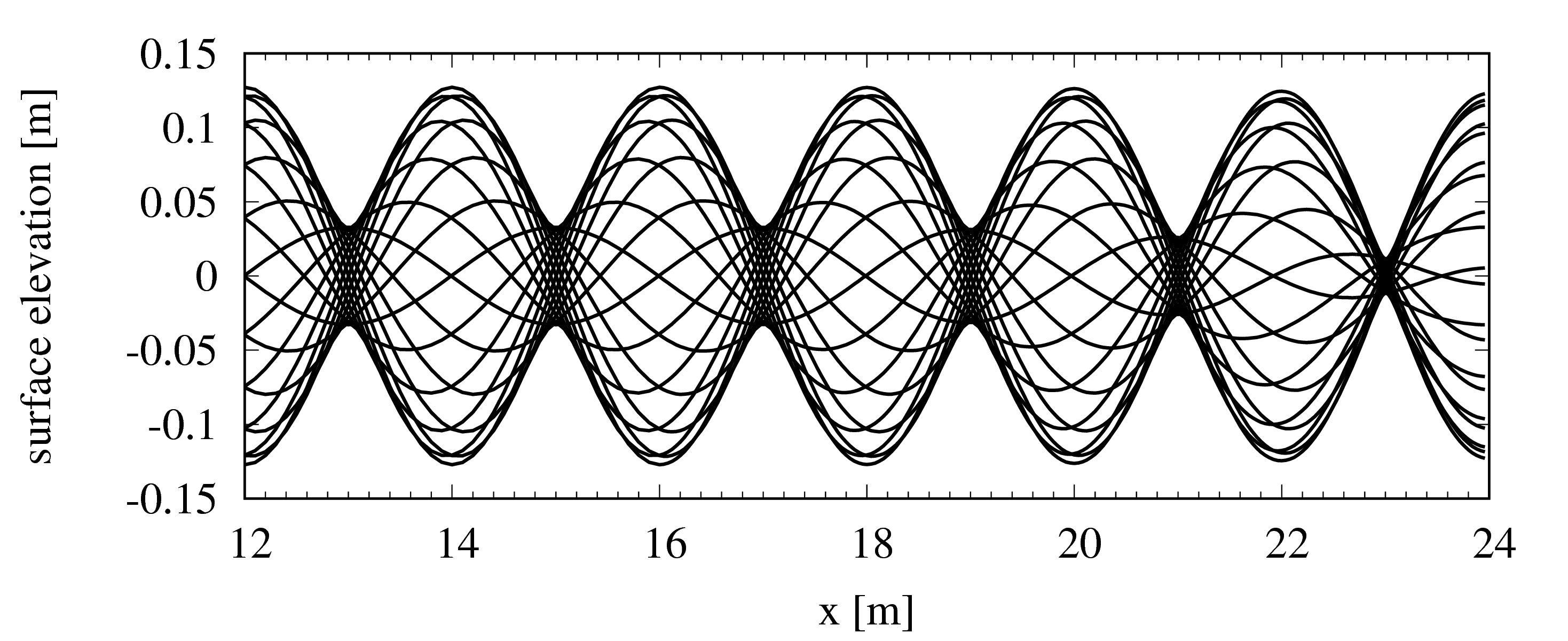}\\
 $\gamma=2.5\, \mathrm{rad/s}$\\
\includegraphics[width=\halffactor\linewidth]{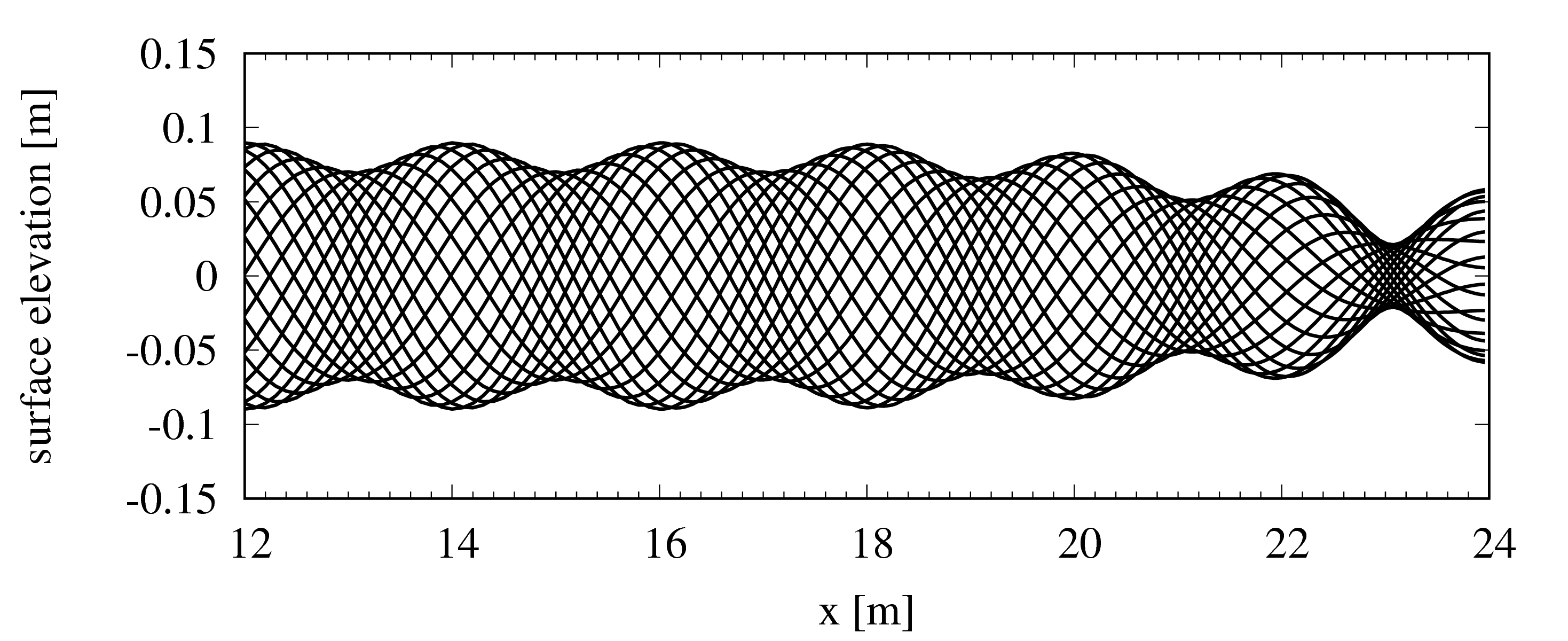}\\
 $\gamma=10\, \mathrm{rad/s}$\\
\includegraphics[width=\halffactor\linewidth]{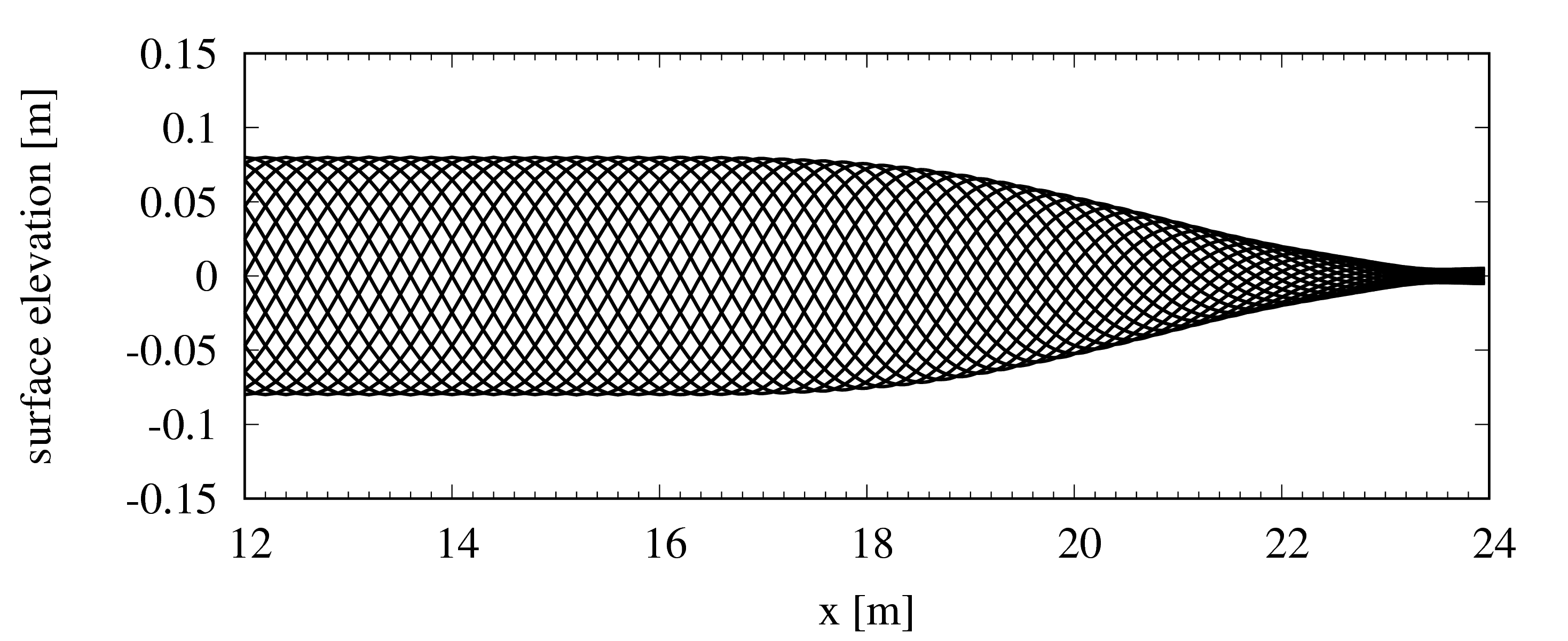}\\
  $\gamma=80\, \mathrm{rad/s}$\\
\includegraphics[width=\halffactor\linewidth]{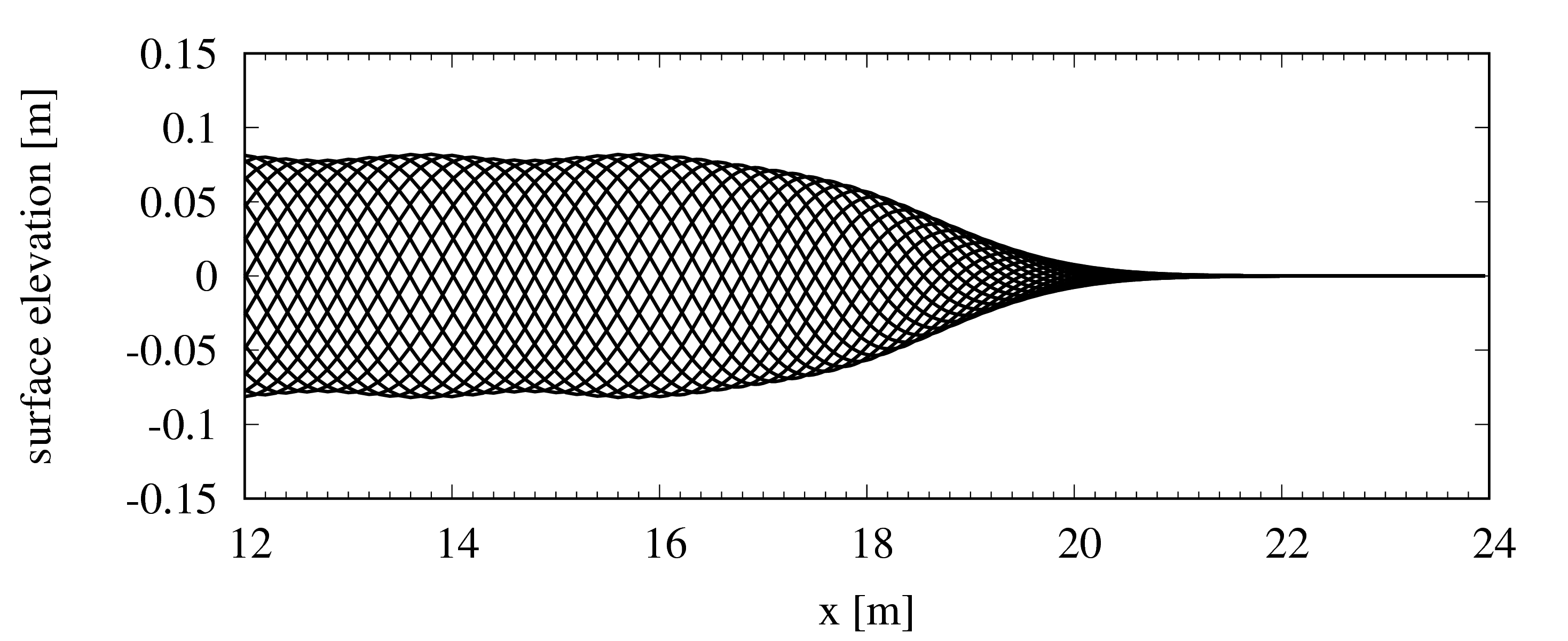}\\
 $\gamma=640\, \mathrm{rad/s}$\\
\includegraphics[width=\halffactor\linewidth]{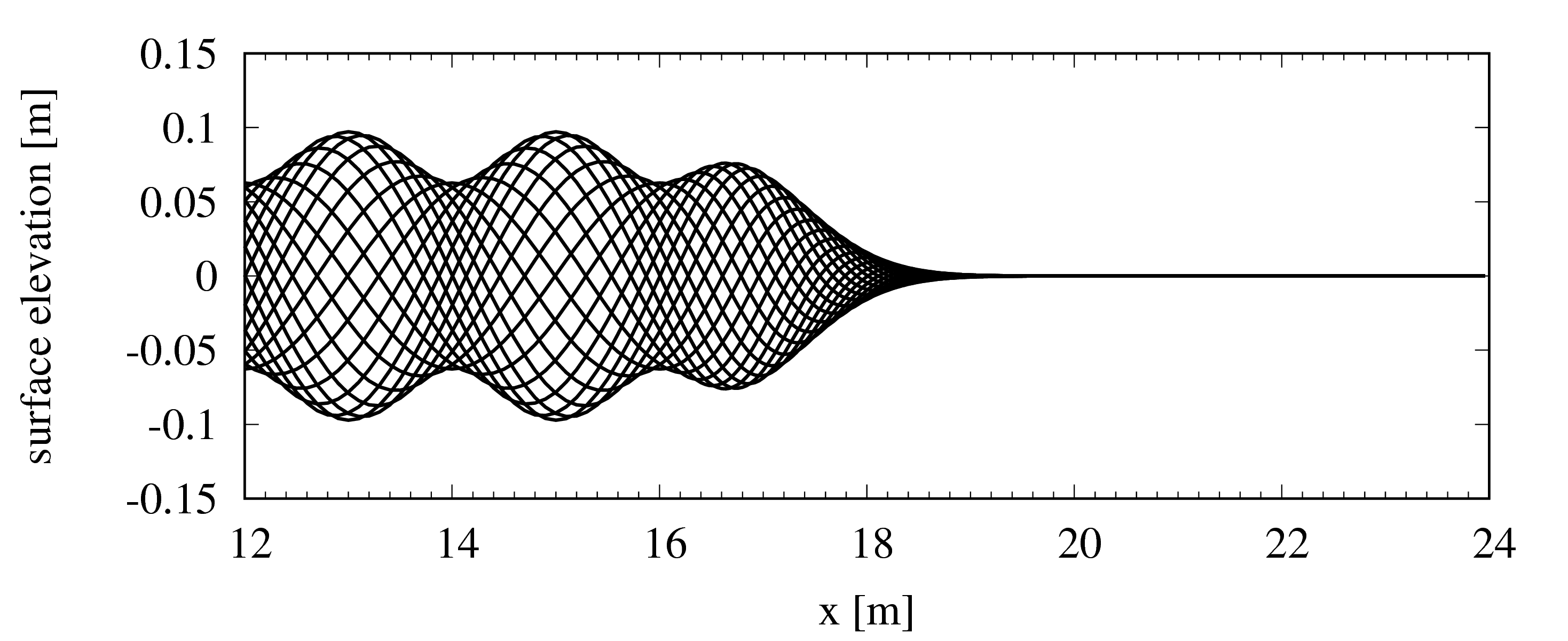}\\
\end{center}
\caption{Theory predictions for surface elevation over $x$-coordinate in the vicinity of a forcing zone with thickness $x_{\mathrm{d}}=2\lambda$ } \label{FIGxelevtheory}
\end{figure}

The simulated and theoretical surface elevations agree well as shown in Figs. \ref{FIGxelevsim} and \ref{FIGxelevtheory}. The peaks in the partial standing wave  have different locations depending on $\gamma$, which shows that, with increasing forcing strength $\gamma$, the effective reflection location shifts from the boundary, to which the forcing zone is attached (here $x=24\, \mathrm{m}$), towards the entrance to the forcing zone (here $x=16\, \mathrm{m}$); this underlines the importance of including reflections which occur within the zone in the analysis.
Just as the surface elevations, the theory predictions for horizontal and vertical velocities agree well with simulation results; since there are no significant differences, these are not plotted here. 

\begin{figure}[H]
\includegraphics[width=\linewidth]{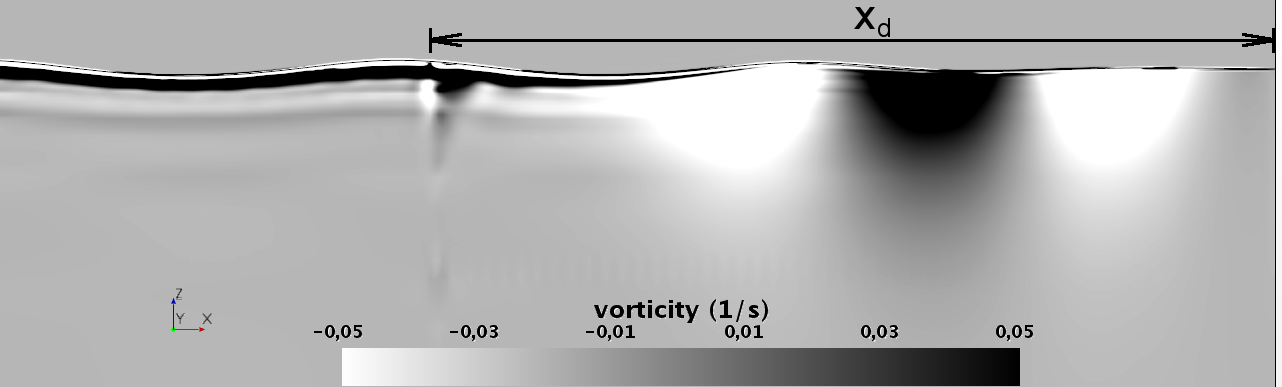}
\caption{Simulation results at time $t=18\, \mathrm{s}$ for vorticity in domain section including the forcing zone, for forcing strength $\gamma=10\, \mathrm{rad/s}$, zone thickness $x_{\mathrm{d}}=2\lambda$, and exponential blending according to Eq. (\ref{EQblendexp})} \label{FIGvortexpsimplot}
\end{figure}

Figure \ref{FIGvortexpsimplot} shows that the forcing produces vorticity (here: $w_{x,j} - u_{z,j}$) in the flow field within the forcing zone.  Although the generated amount of vorticity is comparatively small,  it is clearly visible. 
Further,  small amounts of vorticity are also generated at the free surface and at locations where the mesh size changes, as artifacts of the interface capturing scheme and the discretization; however, such a vorticity generation is discretization dependent, i.e. it disappears on infinitely fine grids, whereas the forcing-generated  vorticity turned out to be grid-independent. 

\begin{figure}[H]
\includegraphics[width=\linewidth]{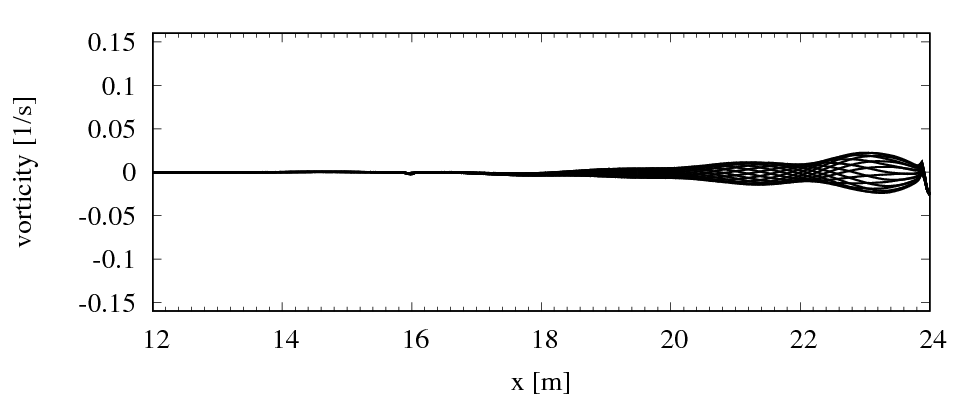}
\includegraphics[width=\linewidth]{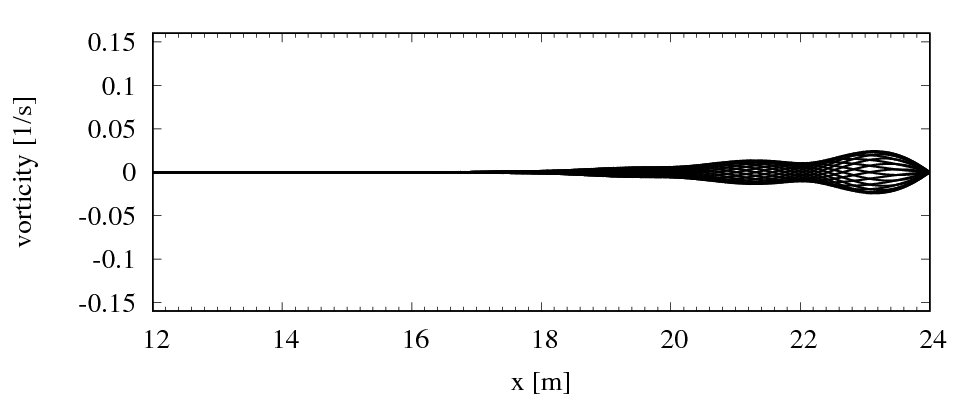}
\caption{Vorticity over $x$-coordinate for simulation (top) and theory (bottom); evaluated at $z=-1.2\, \mathrm{m}$ for equally spaced time intervals during the last simulated period; for lower than optimum forcing strength ($\gamma=2.5\, \mathrm{rad/s}$), zone thickness $x_{\mathrm{d}}=2\lambda$ and exponential blending according to Eq. (\ref{EQblendexp})} \label{FIGvorttheory1}
\end{figure}

\begin{figure}[H]
\includegraphics[width=\linewidth]{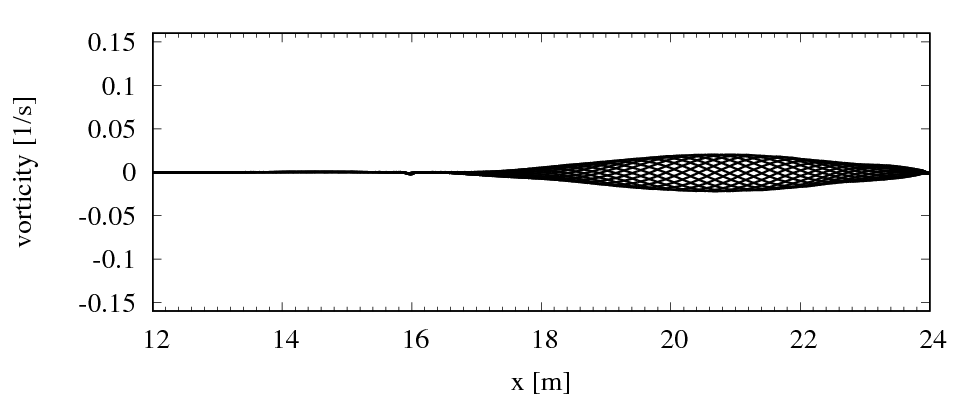}
\includegraphics[width=\linewidth]{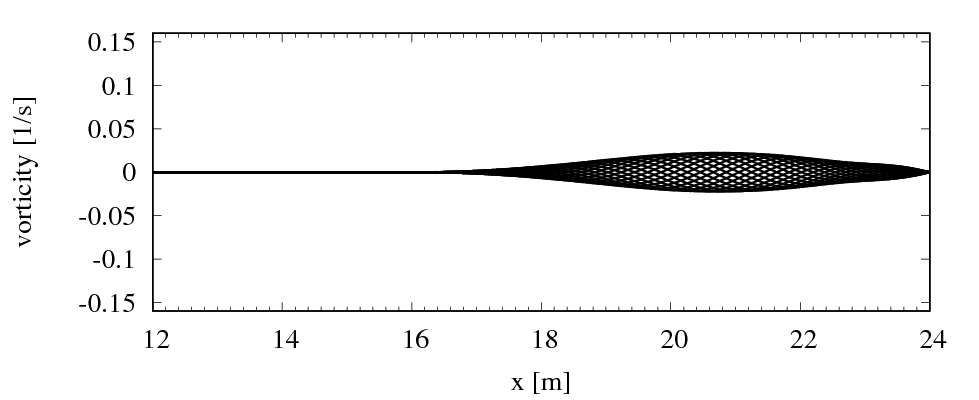}
\caption{As in Fig. \ref{FIGvorttheory1}, except forcing strength close to optimum  ($\gamma=10\, \mathrm{rad/s}$)} \label{FIGvorttheory2}
\end{figure}

\begin{figure}[H]
\includegraphics[width=\linewidth]{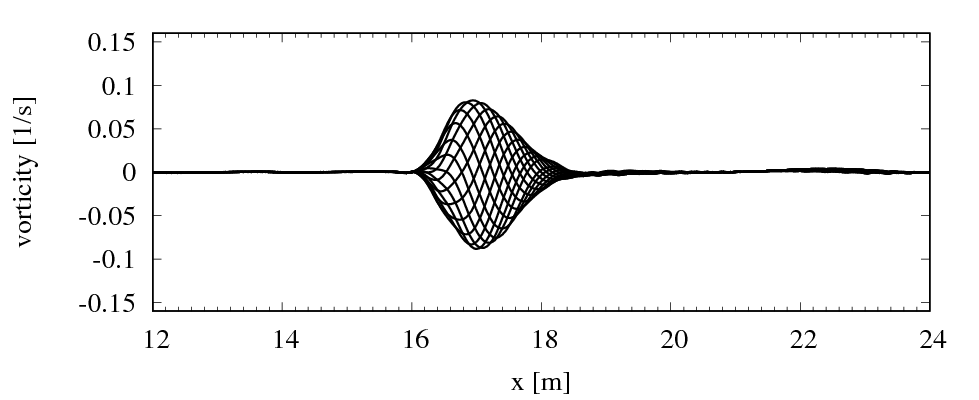}
\includegraphics[width=\linewidth]{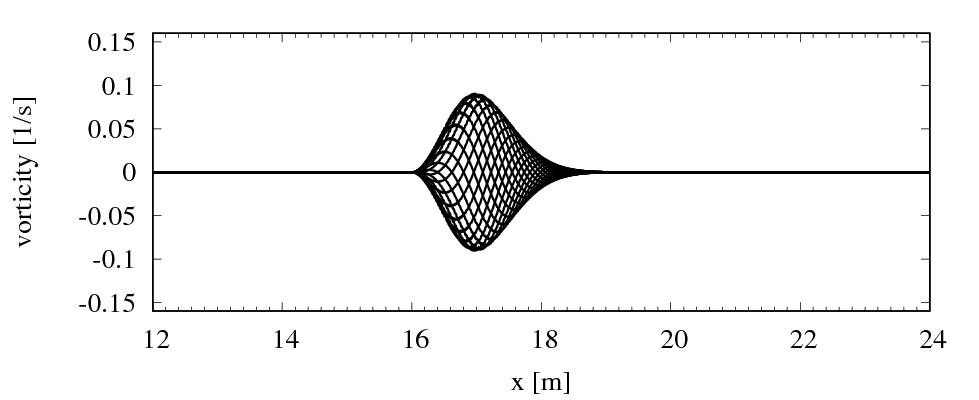}
\caption{As in Fig. \ref{FIGvorttheory1}, except larger than  optimum forcing strength ($\gamma=1280\, \mathrm{rad/s}$)} \label{FIGvorttheory3}
\end{figure}

As Figs. \ref{FIGvorttheory1}, \ref{FIGvorttheory2} and \ref{FIGvorttheory3} show, the theory predicts the forcing-based vorticity generation remarkably well.


 While the velocities and particle displacements are continuous everywhere within the domain according to the theory given in Sect. \ref{SECtheory}, evaluating vorticity $w_{x,j} - u_{z,j}$ for the theory shows that the vorticity is continuous within each zone, as well as at each interface between two zones where the blending function $b(x)$ has the same value. When $b(x)$ changes between two adjacent zones, then the vorticity must be discontinuous at the interface between these zones.

\begin{figure}[H]
\includegraphics[width=\linewidth]{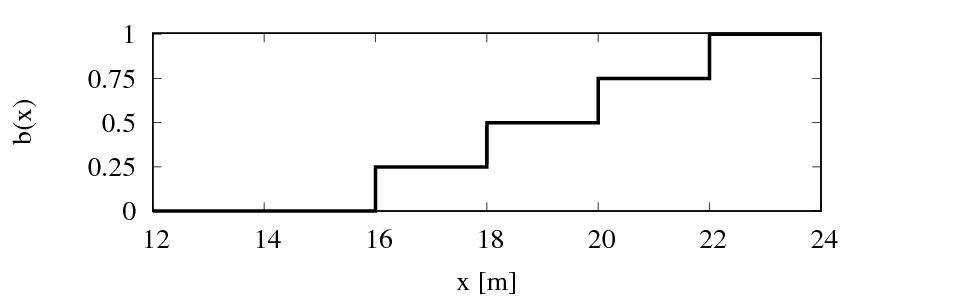}
\caption{Discontinuous blending function $b(x)$ according to Eq. (\ref{EQblendstep}) over $x$-coordinate} \label{FIGvortstepblend}
\end{figure}

\begin{figure}[H]
\includegraphics[width=\linewidth]{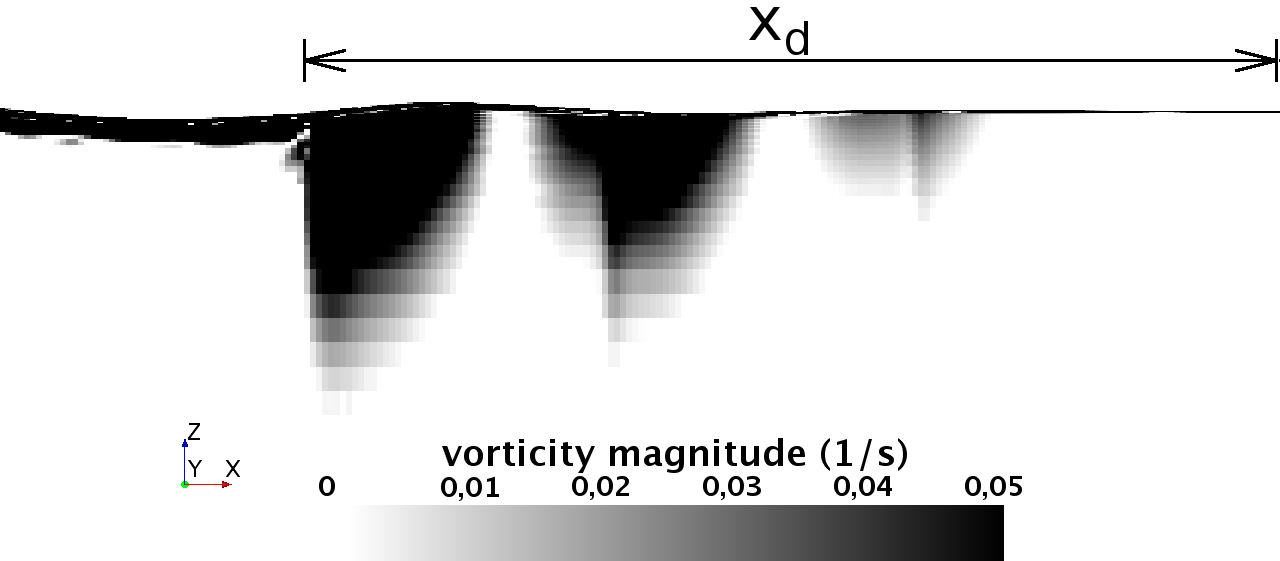}
\caption{Simulation results at time $t\approx 17.06\, \mathrm{s}$ for vorticity magnitude in domain section including the forcing zone, for forcing strength $\gamma=10\, \mathrm{rad/s}$, zone thickness $x_{\mathrm{d}}=8\, \mathrm{m}$, and step-like blending according to Eq. (\ref{EQblendstep})} \label{FIGvortstepsim}
\end{figure}

\begin{figure}[H]
\includegraphics[width=\linewidth]{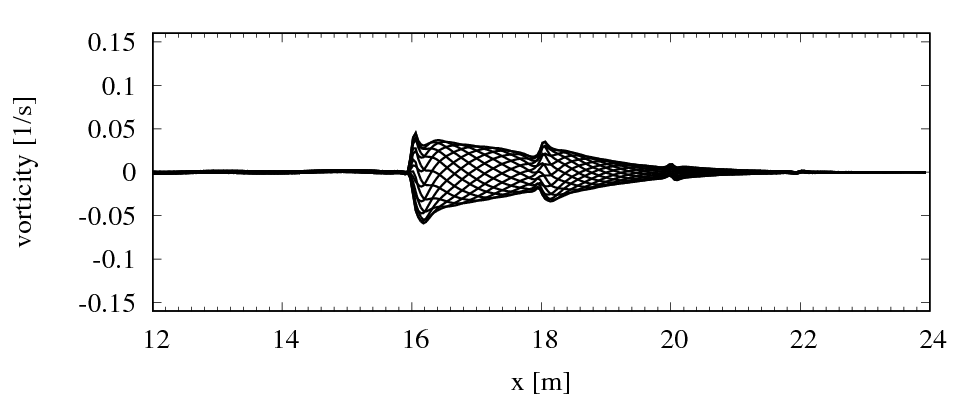}
\includegraphics[width=\linewidth]{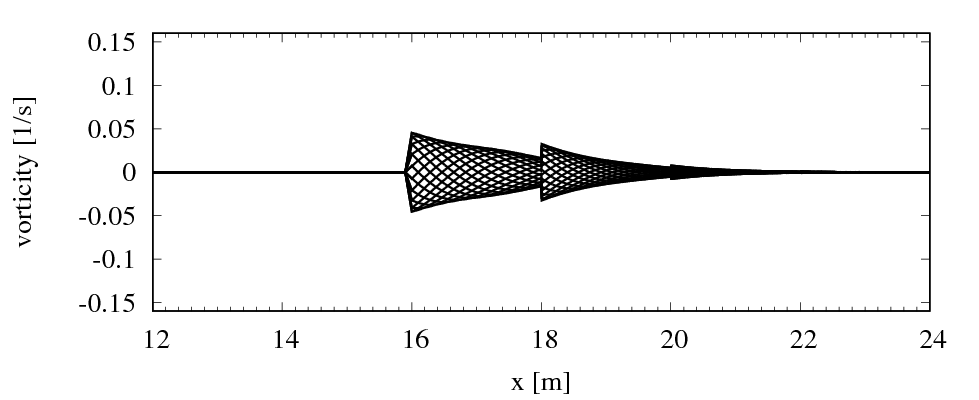}
\caption{Vorticity over $x$-coordinate for simulation (top) and theory (bottom); evaluated at depth $z=-1.2\, \mathrm{m}$ for equally spaced time intervals during the last simulated period; for forcing strength $\gamma=10\, \mathrm{rad/s}$, zone thickness $x_{\mathrm{d}}=8\, \mathrm{m}$ and step-like blending according to Eq. (\ref{EQblendstep})} \label{FIGvortsteptheory}
\end{figure}

To show that this is realistic and occurs in the flow simulations as well, the simulations are rerun using a strongly discontinuous blending function

\begin{equation}
b(x) =
  \begin{cases}
    0.25       & \quad \text{if } 16\, \mathrm{m} \leq x <18\, \mathrm{m} \\
    0.5       & \quad \text{if } 18\, \mathrm{m} \leq x <20\, \mathrm{m} \\
    0.75       & \quad \text{if } 20\, \mathrm{m} \leq x <22\, \mathrm{m} \\
    1.0       & \quad \text{if } 22\, \mathrm{m} \leq x \leq 24\, \mathrm{m} \\
    0       & \quad \text{else} \quad ,
  \end{cases}
\label{EQblendstep}
\end{equation}
with forcing zone thickness $x_{\mathrm{d}} = 24\, \mathrm{m} - 16\, \mathrm{m} =8\, \mathrm{m}\approx 2\lambda$; $b(x)$ is illustrated in Fig. \ref{FIGvortstepblend}.
Figures \ref{FIGvortstepsim} and \ref{FIGvortsteptheory} show that also in the flow simulations the vorticity  is discontinuous between cells where $b(x)$ changes, and is continuous otherwise; the agreement between theory and simulation was again satisfactory, as is exemplarily shown for  $\gamma=10\, \mathrm{rad/s}$ in Fig. \ref{FIGvortsteptheory}.

\subsection{Forcing of $z$-momentum}
\label{SECresz}
\begin{figure}[H]
\includegraphics[width=\linewidth]{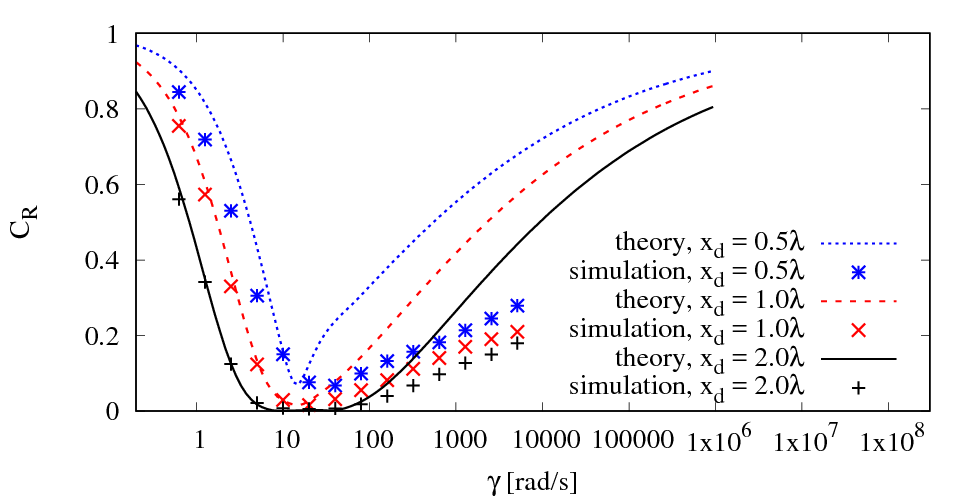} 
\caption{Reflection coefficient $C_{\mathrm{R}}$ over forcing strength $\gamma$ for forcing of $z$-momentum, given for different forcing zone thickness $x_{\mathrm{d}} $  according to simulation and theory } \label{FIGCrGammaZmom}
\end{figure}

\begin{figure}[H]
\begin{center}
 $\gamma=2.5\, \mathrm{rad/s}$\\
\includegraphics[width=\halffactor\linewidth]{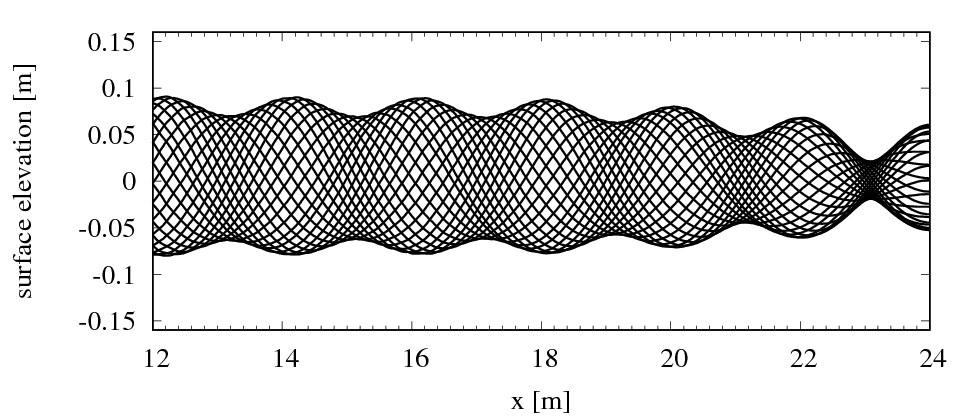}\\
 $\gamma=10\, \mathrm{rad/s}$\\
\includegraphics[width=\halffactor\linewidth]{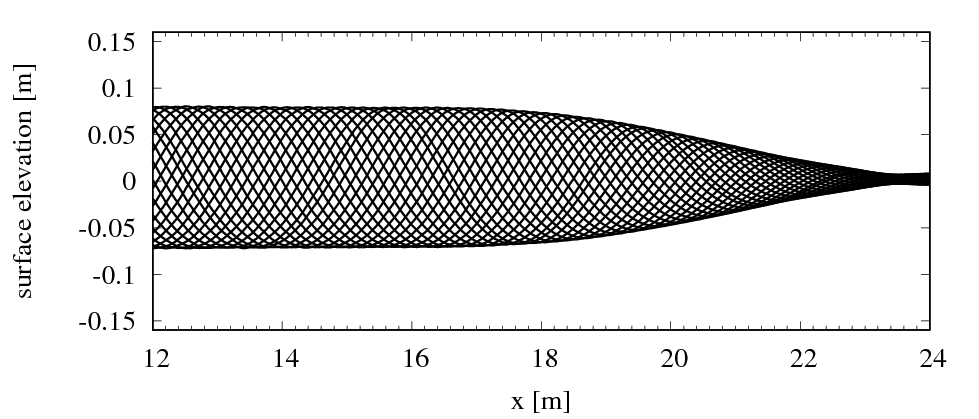}\\
 $\gamma=640\, \mathrm{rad/s}$\\
 \includegraphics[width=\halffactor\linewidth]{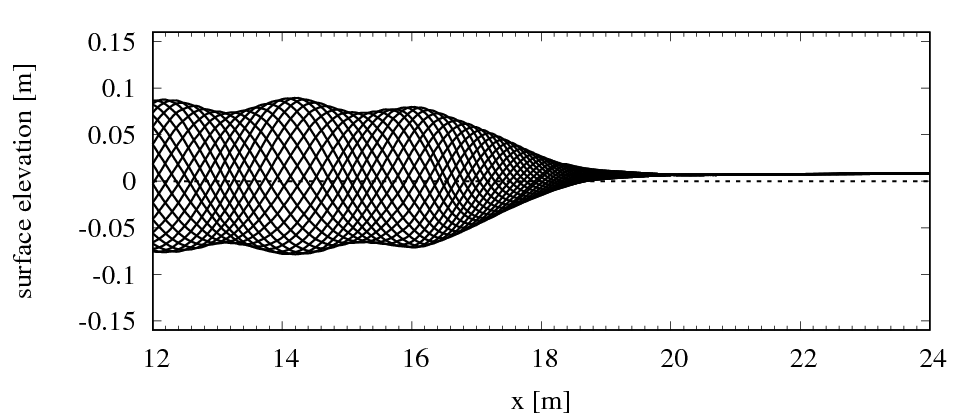}\\
 $\gamma=2560\, \mathrm{rad/s}$\\
\includegraphics[width=\halffactor\linewidth]{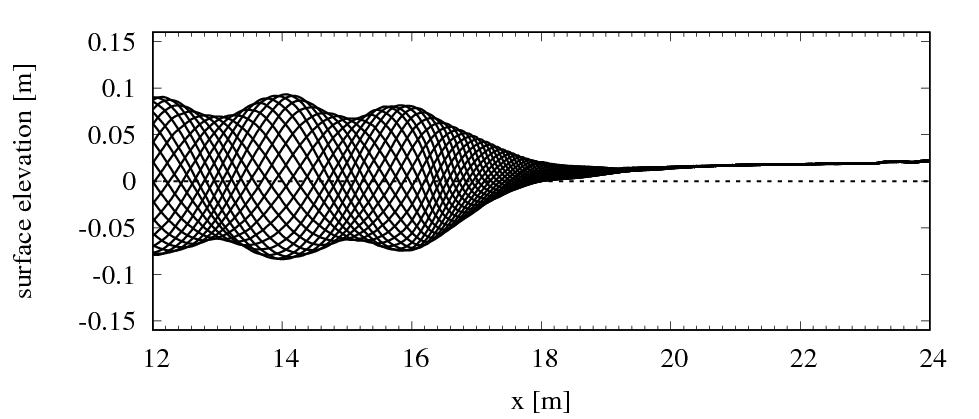}\\
\end{center}
\caption{Simulation results for surface elevation over $x$-coordinate in the vicinity of a forcing zone with thickness $x_{\mathrm{d}}=2\lambda$ and forcing of $z$-momentum; evaluated for equally spaced time intervals during the last simulated period; for $\gamma\lesssim $ than optimum (upper two plots) there are no noticeable differences to $x$-momentum forcing; for stronger than optimum damping (lower two plots) a rise in water level (dashed line indicates undisturbed free surface) occurs close to the domain boundary at $x=24\, \mathrm{m}$, which indicates a net mass flux into the forcing zone that does not occur for $x$-momentum forcing} \label{FIGzelevsim}
\end{figure}

Repeating the simulations from Sect. \ref{SECresx} with forcing of $z$-momentum instead of $x$-momentum shows that the theory predicts optimum forcing strength and corresponding reflection coefficient $C_{\mathrm{R}}$ reliably. 
Comparing Figs. \ref{FIGCrGammaZmom} and \ref{FIGzelevsim} with Figs. \ref{FIGCrGammaXmomLOG} to \ref{FIGxelevtheory} shows that, for optimum or lower values of forcing strength $\gamma$, the results for forcing of $x$- and $z$-momentum both agree well with theory predictions.

For stronger forcing than optimum (here: $\gamma>40\, \mathrm{rad/s}$) the theory predictions of $C_{\mathrm{R}}$ are conservative. Figure \ref{FIGzelevsim} shows that for such $\gamma$-values the mean surface elevation increases within the forcing zone. This becomes more pronounced with increasing $\gamma$. Thus stronger-than-optimum forcing of $z$-momentum leads to a noticeable net mass flux into the forcing zone in the simulation.
In the present case, this mass flux resulted in a lower reflection coefficient $C_{\mathrm{R}}$ compared with $x$-momentum forcing for the same $\gamma$.

However, since this effect is negligible for optimum and lower forcing strength,  the theoretical prediction is  satisfactory for practical purposes. 

\subsection{Forcing of $x$- and $z$-momentum}
\label{SECresxz}
\begin{figure}[H]
\includegraphics[width=\linewidth]{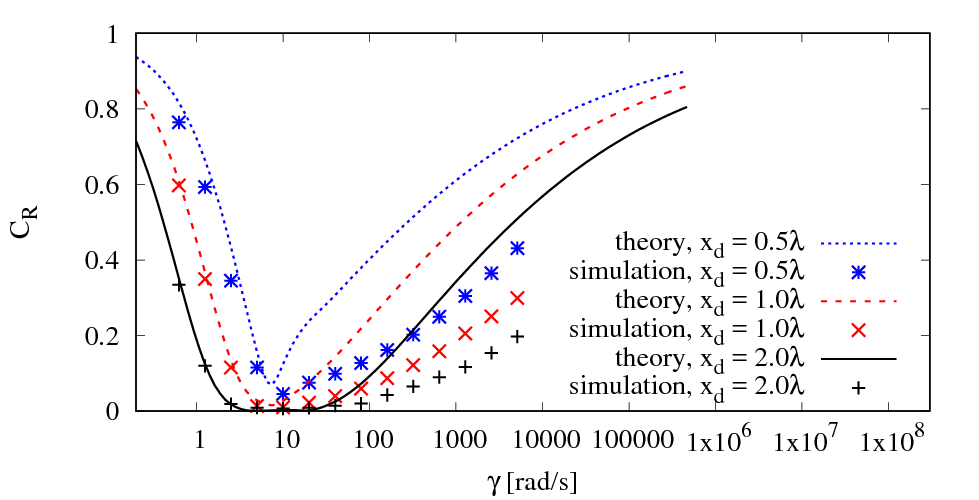} 
\caption{Reflection coefficient $C_{\mathrm{R}}$ over forcing strength $\gamma$ for forcing of $x$- and $z$-momentum, given for different forcing zone thickness $x_{\mathrm{d}} $  according to simulation and theory } \label{FIGCrGammaXZmom}
\end{figure}

Repeating the simulations from Sect. \ref{SECresx} with forcing of both $x$- and $z$-momentum shows that the theory, with the extension given in Sect. \ref{SECtheoryxza}, reliably predicts the optimum forcing strength $\gamma $ and the corresponding reflection coefficient $C_{\mathrm{R}}$. As expected from the results in Sects. \ref{SECresx} and \ref{SECresz}, Fig. \ref{FIGCrGammaXZmom} shows that for stronger than optimum forcing the theory overpredicts the reflection coefficients.

\subsection{Forcing of volume fraction $\alpha$ and $x$- and $z$-momentum}
\label{SECresaxz}
\begin{figure}[H]
\includegraphics[width=\linewidth]{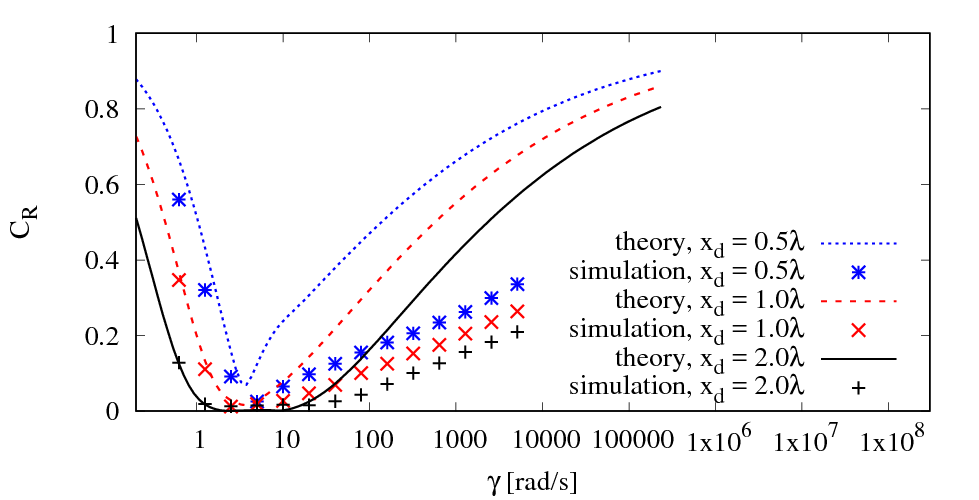} 
\caption{Reflection coefficient $C_{\mathrm{R}}$ over forcing strength $\gamma$ for forcing of $x$- and $z$-momentum and volume fraction $\alpha$, given for different forcing zone thickness $x_{\mathrm{d}} $  according to simulation and theory } \label{FIGCrGammaXZAmom}
\end{figure}

Repeating the simulations from Sect. \ref{SECresx} with forcing of volume fraction $\alpha$ and both $x$- and $z$-momentum shows that the theory, with the extension given in Sect. \ref{SECtheoryxza}, reliably predicts the optimum forcing strength and the corresponding reflection coefficient. Figure \ref{FIGCrGammaXZAmom} shows that for stronger than optimum damping the theory overpredicts the reflection coefficients, otherwise the results agree well.

\section{Conclusion}
\label{SECconcl}
To reliably damp surface waves using forcing-zone-type approaches (such as absorbing layers, sponge layers, relaxation zones, etc.), the case-dependent parameters in the source functions must be adjusted to the wave. These adjustments are not 'ad-hoc'. Instead, there is a mathematical foundation for choosing the tunable parameters.
The parameter forcing strength $\gamma$ scales with the angular wave frequency $\omega$, and the layer thickness $x_{\mathrm{d}}$ scales with wavelength $\lambda$.

In this work, a theory was presented which  reliably predicts the  reflection coefficients for forcing zones with sufficient accuracy for practical purposes. The theory applies for forcing of horizontal fluid velocities, forcing of vertical fluid velocities, forcing of the volume fraction, as well as for combinations of these approaches.
The theory is given for a general forcing source term formulation, so that it can be easily applied to many existing forcing-zone-type approaches from literature and  existing implementations in commercial flow solvers. A computer code for evaluating the theory has been made available to the scientific community as free software.

The main benefit of the theory is that it can be used to optimally select the forcing zone parameters before running the simulation.  The present results illustrate the importance of adjusting these parameters for every flow simulation, since the common practice to use default settings can produce large errors. 

The theory predictions for forcing of horizontal velocities were remarkably accurate both for reflection coefficients and the flow within the layer. This also held for forcing of vertical velocities and volume fraction, except for the case when the forcing strength was larger than optimum; then, the reflection was lower than predicted by the theory. This was attributed to a net mass flux into the forcing layer, which the theory does not predict. 
However, for practical purposes the theory works satisfactory for all forcing types discussed, since the optimum parameters and the corresponding reflection coefficients are predicted reliably, and otherwise the predictions are conservative.

Apart from predicting the reflection before the simulation is run, the theory can also be used to assess the amount of undesired wave reflections after the simulation is run. This is especially useful for classification societies, when they have to assess the quality of results from flow simulations that they did not perform themselves. Since it is not industry standard to provide reflection coefficients for flow simulations, it is beneficial to be able to quickly take any forcing zone formulation that can be expressed in terms of Eqs. (\ref{EQmomdamp}) and (\ref{EQmassdamp}), obtain the reflection coefficients from theory, and judge whether the settings were appropriate.

Future research is necessary to verify how accurate the theory predicts the damping of three-dimensional waves, irregular waves and highly non-linear waves, such as rogue waves, waves close to breaking steepness and even breaking waves; literature results are promising that the theory covers these cases as well (Peri\'c et al., 2015; Peri\'c and Abdel-Maksoud, 2016). 
Further, research is necessary regarding the optimum choice of blending function. Moreover, applying and possibly extending the theory to the simultaneous generation and absorption of waves via forcing zones, and subsequently the coupling of different flow solvers (e.g. viscous and inviscid codes), seems promising in the light of the present findings. 

\appendix
\section{Forcing proportional to the velocity squared}
\label{SECappendixA}
Examples of widely used wave damping implementations are the approaches by Choi and Yoon (2009, implemented in STAR-CCM+ by Siemens) and by Park et al. (1999, implemented in ANSYS Fluent), which are
\begin{equation}
q_{z_{\mathrm{CCM+}}} = \underbrace{  \gamma b(x) (-w) }_{\mathrm{forcing}\ \propto \ w} + \underbrace{  \tilde{\gamma} b(x) (-w) |w|}_{\mathrm{forcing}\ \propto \ w|w|}\quad ,
\label{EQdampchoi}
\end{equation}
\begin{equation}
q_{z_{\mathrm{ANSYS}}}  =  \underbrace{0.5 \left(1-\frac{z-z_{{\rm fs}}}{z_{{\rm b}}-z_{{\rm fs}}}\right)}_{\mathrm{additional\ factor}} \underbrace{\tilde{\gamma} b(x) (-w) |w|}_{\mathrm{forcing}\ \propto \ w|w|} \quad ,
\label{EQdampANSYS}
\end{equation}
with vertical velocity $w$, forcing strengths $\gamma$ and $\tilde{\gamma}$, vertical coordinate $z$ with domain bottom at $z_{{\rm b}}$, and vertical location $z_{{\rm fs}}$ of the free surface.  
Further $b(x)$ is exponential (Eq. (\ref{EQblendexp})) in Eq. (\ref{EQdampchoi}) and quadratic (Eq. (\ref{EQblendquad})) in Eq. (\ref{EQdampANSYS}).
In Eq. (\ref{EQdampchoi}),  the first term corresponds to Eq. (\ref{EQmomdamp}) and is therefore directly proportional to the vertical velocity, while the second term contains an additional factor $|w|$, which renders this forcing term directly proportional to $w|w|$. 
Equation (\ref{EQdampANSYS}) corresponds to the second term in Eq. (\ref{EQdampchoi}), except for a factor $0.5$, an additional vertical blending $\left(1-\frac{z-z_{{\rm fs}}}{z_{{\rm b}}-z_{{\rm fs}}}\right)$ and a slightly different $b(x)$ (see Fig. \ref{FIGblend}).

At the time of writing, the default values in the commercial codes for forcing strengths in Eqs. (\ref{EQdampchoi}) and (\ref{EQdampANSYS}) are  $ \gamma = 10.0  \, \mathrm{rad/s}$ and  $ \tilde{\gamma} = 10.0 \, \mathrm{m^{-1}}$. Peri\'c and Abdel-Maksoud (2016) found that for a $w|w|$-proportional forcing, as in the second term in Eq. (\ref{EQdampchoi}) and in Eq. (\ref{EQdampANSYS}), the optimum value for $\tilde{\gamma}$ is more than one order of magnitude larger than the optimum value for  $\gamma$. Thus with default settings in STAR-CCM+, the second term in Eq. (\ref{EQdampchoi}) has a negligible effect compared to the first term. 

\begin{figure}[H]
\begin{center}
\includegraphics[width=\linewidth]{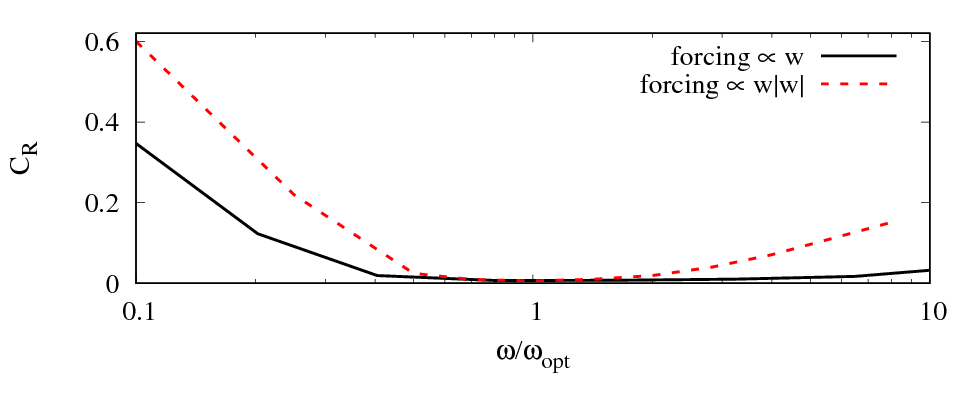}
\end{center}
\caption{Reflection coefficient $C_{\mathrm{R}}$ over angular wave frequency $\omega$ scaled by optimum angular wave frequency $\omega_{\mathrm{opt}}$, given for simulation results from Peri\'c and Abdel-Maksoud (2016); for forcing zone thickness $x_{\mathrm{d}}=2\lambda$ and blending $b(x)$ according to Eq. (\ref{EQblendexp}); 
for forcing using only the first term in Eq. (\ref{EQdampchoi}) (continuous line) with fixed forcing strength $\gamma$, versus using only the second term in Eq. (\ref{EQdampchoi}) (dotted line) with fixed forcing strength  $\tilde{\gamma}$
} \label{FIGCr0omega}
\end{figure}

Both $w$-proportional and $w|w|$-proportional forcing produced comparable reflection coefficients at optimum settings, so both approaches can be used  to damp waves successfully. However, for a fixed forcing strength,  directly-proportional forcing as in Eq. (\ref{EQmomdamp})  has a wider range of wave frequencies which are damped satisfactorily, as illustrated in Fig. \ref{FIGCr0omega}. Historically, $w|w|$-proportional forcing terms may have been introduced as analogy to porous media flows, where for larger flow rates effects like turbulence lead to nonlinearities which can be expressed as quadratically dependent on the flow velocity, such as the Forchheimer or Brinkman extension to Darcy's law, see Straughan (2008). However, this analogy is not entirely valid.
Even in steep nonlinear ocean waves, turbulence effects are insignificant unless there is wave breaking, which especially with regard to coupling of different flow solvers should not be provoked inside the forcing zone. Moreover, Fourier approximation methods allow to split nonlinear waves into different regular harmonics, so applying a forcing directly proportional to the velocity according to Eq. (\ref{EQmomdamp}) acts on the higher harmonics as well, so the damping of the higher harmonics is already accounted for in $w$-proportional forcing.

\section{Theory implications for tuning forcing strength $\gamma$ and for the choice of blending function $b(x)$}
\label{SECappendixB}
Since the theory from Sect. \ref{SECtheory} was validated in Sect. \ref{SECresdisc} and can be considered sufficiently accurate for practical purposes, the following sections show theoretical predictions without backup from simulation results. This is mainly because, to obtain the following results,  a very large  number of simulations would be required, the combined computational effort being out of the scope of this study. 

\begin{figure}[H]
\includegraphics[width=\linewidth]{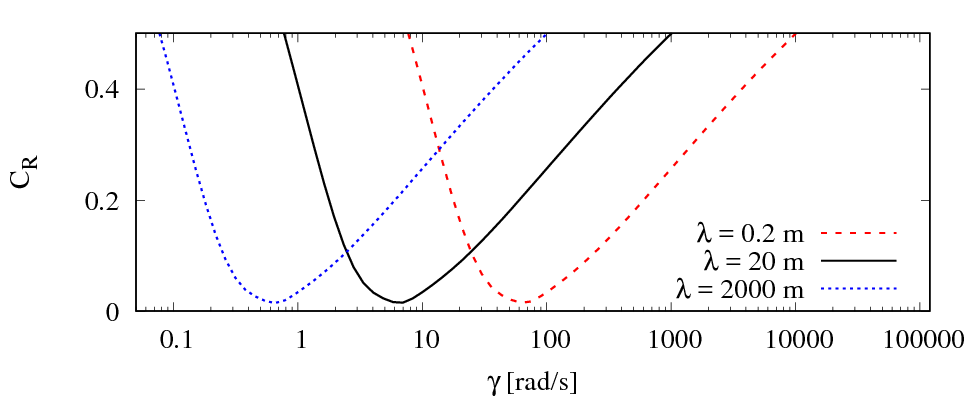} 
\caption{Theory prediction for reflection coefficient $C_{\mathrm{R}}$ over forcing strength $\gamma$; for forcing of $x$-momentum with zone thickness $x_{\mathrm{d}} = 1\lambda $, $b(x)$ from Eq. (\ref{EQblendexp}) and deep-water waves; for three different wavelengths $\lambda=0.2\, \mathrm{m}, 20\, \mathrm{m}, 2000\, \mathrm{m}$; it is evident that forcing strength $\gamma$ must be adjusted for each wave } \label{FIGdifferentlambda}
\end{figure}

For a typical forcing zone setup, Fig. \ref{FIGdifferentlambda} illustrates  the necessity of adjusting the forcing strength $\gamma$ for different waves.

\begin{figure}[H]
\includegraphics[width=\halffactor\linewidth]{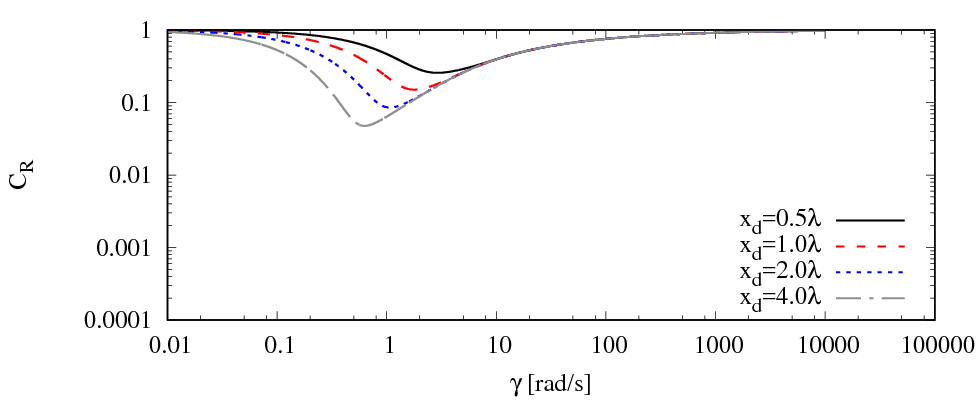} \\
\includegraphics[width=\halffactor\linewidth]{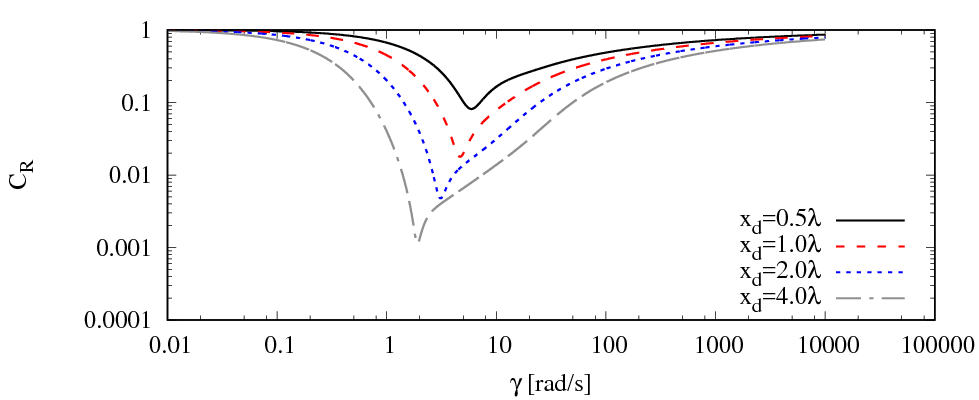} \\
\includegraphics[width=\halffactor\linewidth]{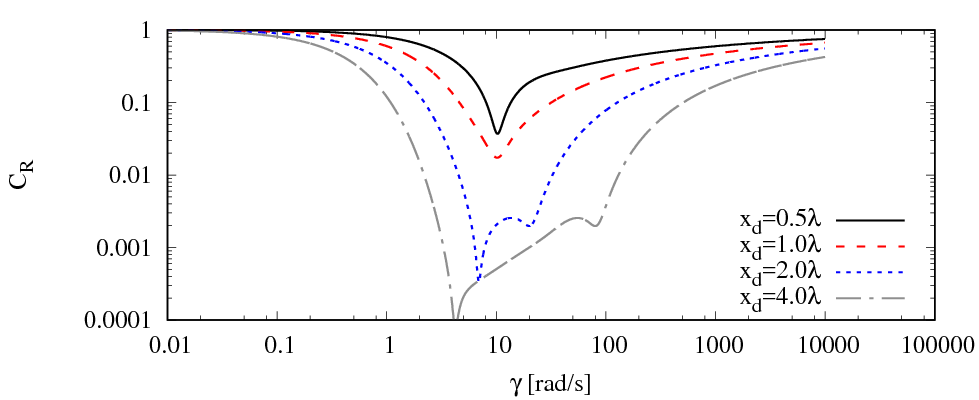} \\
\includegraphics[width=\halffactor\linewidth]{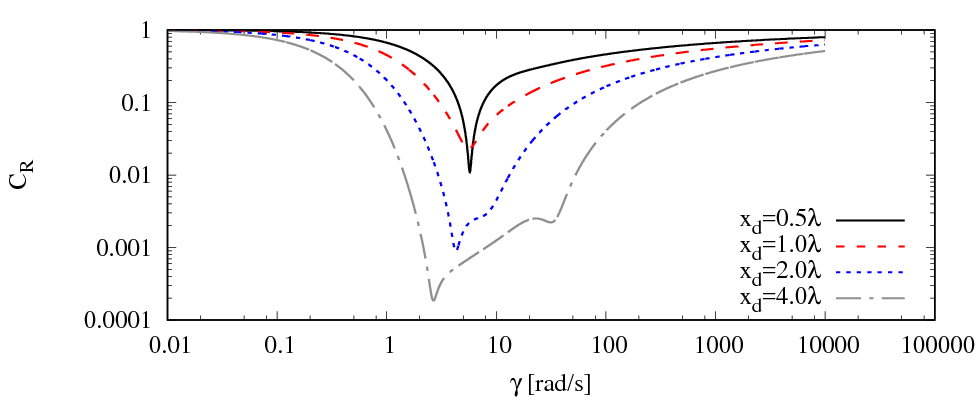} \\
\includegraphics[width=\halffactor\linewidth]{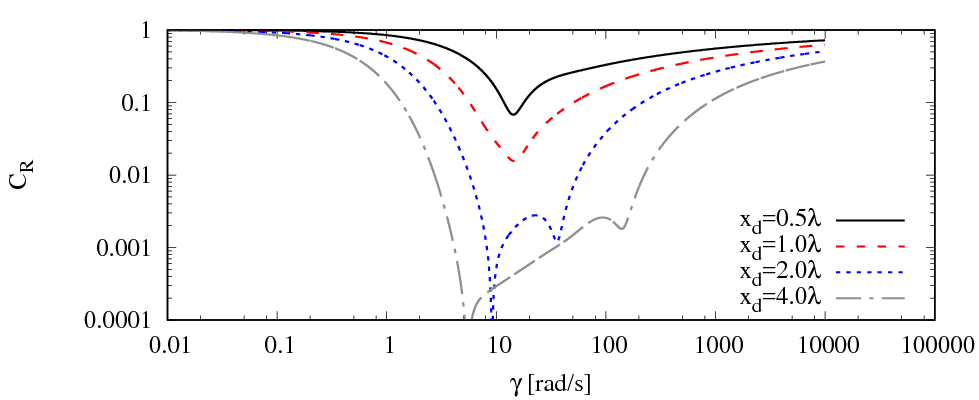} 
\caption{Theory prediction for reflection coefficient $C_{\mathrm{R}}$ over forcing strength $\gamma$; for forcing of $x$-momentum with different forcing zone thickness $0.5\lambda \leq x_{\mathrm{d}} \leq 4\lambda $ and deep-water waves with period $T=1.6\, \mathrm{s}$; from top to bottom: constant (Eq. (\ref{EQblendconst})), linear (Eq. (\ref{EQblendlin})), quadratic (Eq. (\ref{EQblendquad})), cosine-square (Eq. (\ref{EQblendcos2})) and exponential (Eq. (\ref{EQblendexp})) blending } \label{FIGdifferentbx}
\end{figure}

So far, the optimum choice for the blending term $b(x)$ for a given zone thickness $x_{\mathrm{d}}$ is not known. However, Fig. \ref{FIGdifferentbx} shows that common choices for blending functions such as quadratic blending (Eq. (\ref{EQblendquad})), cosine-square blending (Eq. (\ref{EQblendcos2})) and exponential blending according to Eq. (\ref{EQblendexp}) perform similarly well, with perhaps the exponential blending performing slightly better.
Thus these blending functions can all be used successfully. 

Constant blending (Eq. (\ref{EQblendconst})) and linear blending (Eq. (\ref{EQblendlin})) are not recommended, since they lead to considerably higher reflection coefficients.

\section{Convergence of theory to solution for continuous blending}
\label{SECappendixC}

\begin{figure}[H]
\includegraphics[width=\linewidth]{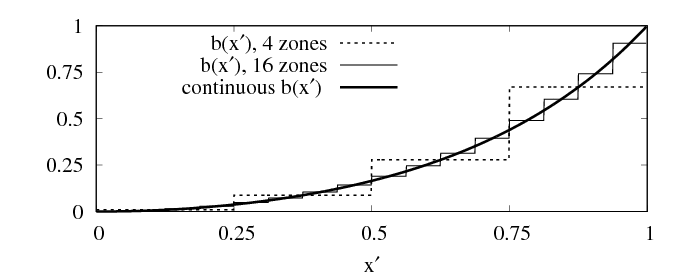} 
\caption{Blending function $b(x')$ according to Eq. (\ref{EQblendexp}) over $x'$-coordinate; $x'$ is directed in wave propagation direction and linearly scaled such that it is $0$ at the entrance to the forcing zone and $1$ at the boundary to which the zone is attached; for forcing layer consisting of $4$ zones, $16$ zones and for continuous $b(x')$ } \label{FIGdiscBcomp}
\end{figure}

The theory from Sect. \ref{SECtheory} subdivides the forcing zone into $n$ zones with constant blending $b(x)$ as illustrated in Fig. \ref{FIGdiscBcomp}. 
This section demonstrates that, if $n$ is larger than a certain threshold, then the theory results can be considered independent of $n$. Thus also the wave damping in flow simulations is basically grid-independent, if the number $n$ of grid cells, by which the forcing zone is discretized in wave propagation direction, is above the same threshold.

\begin{figure}[H]
\includegraphics[width=\linewidth]{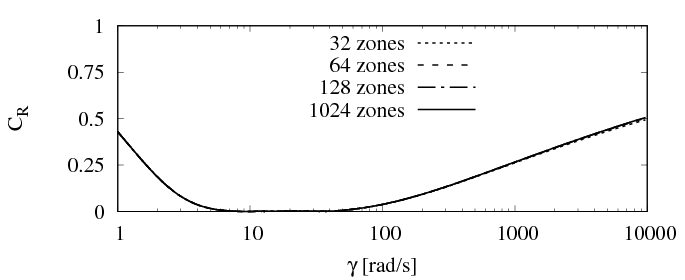} 
\includegraphics[width=\linewidth]{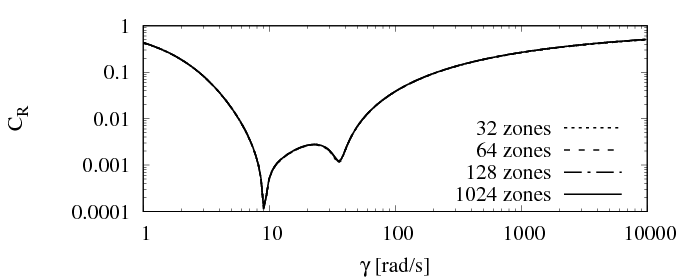} 
\caption{Theory prediction for reflection coefficient $C_{\mathrm{R}}$ over forcing strength $\gamma$; for forcing of $x$-momentum with zone thickness $x_{\mathrm{d}} = 2\lambda $, wave period $T=1.6\, \mathrm{s}$, deep-water conditions, and $b(x)$ from Eq. (\ref{EQblendexp}); for   $b(x)$ subdivided into $32$, $64$, $128$ and $1024$ zones } \label{FIGCr0gdisc}
\end{figure}

Figure \ref{FIGCr0gdisc}  shows for a subdivision into $32$ zones, that the results are barely distinguishable from subdivisions into larger numbers of zones. This agrees well with findings by Peri\'c and Abdel-Maksoud (2016), where the wave damping was observed to be grid-independent for practical flow simulation setups (i.e. grids with at least $30$ cells per wavelength). 

\begin{figure}[H]
\includegraphics[width=\linewidth]{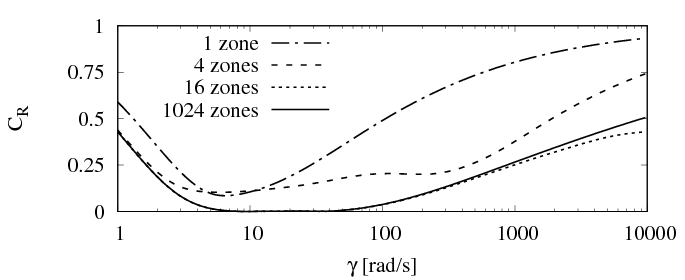} 
\includegraphics[width=\linewidth]{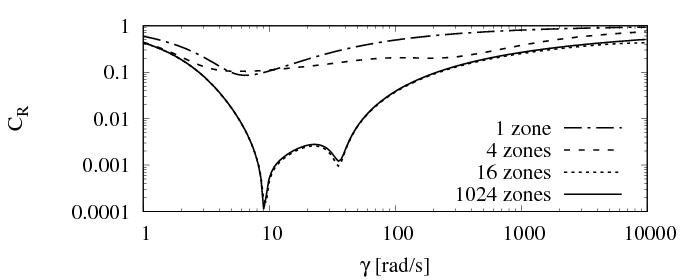} 
\caption{As Fig. \ref{FIGCr0gdisc}, except that $b(x)$ is subdivided into $1$, $4$, $16$ and $1024$ zones} \label{FIGCr0gdiscLOG}
\end{figure}

Figure \ref{FIGCr0gdiscLOG} shows that for subdivision into less than $16$ zones, the results differ significantly from the results for $>32$ zones. This is relevant when assessing flow simulations, in which a combination of  grid stretching and forcing zones is used to damp the waves; in industrial practice, these two wave damping approaches are sometimes combined with the intention to lower the computational effort and to improve the damping. However, Figs. \ref{FIGCr0gdisc} and \ref{FIGCr0gdiscLOG} show that, if due to the grid stretching the number of grid cells per zone thickness drops below a certain threshold, then grid stretching can significantly increase reflection coefficient $C_{\mathrm{R}}$.  Based on the present results, it is recommended to have cell sizes of at least $\lambda/10$ when combining grid stretching and forcing zones. 

If the forcing zone is subdivided into a sufficient number of zones $n$, then the difference between the theory solutions for different $n$  can be estimated by a Richardson-type extrapolation. Detailed information on Richardson extrapolation can be found e.g. in Richardson (1911), Richardson and Gaunt (1927), and Ferziger and Peri\'c (2002).
Say
\begin{equation}
A = A_{\mathrm{h}} + \varepsilon_{\mathrm{h}} \quad ,
\label{EQAh}
\end{equation}
where $A$ is the analytical solution for $n=\infty$, $A_{\mathrm{h}}$ is the analytical solution for $n=x_{\mathrm{d}}/h$, and $\varepsilon_{\mathrm{h}}$ is the error. Let all zones have the same thickness $x_{\mathrm{d},j}=h$, with a total forcing zone thickness of $x_{\mathrm{d}}=\sum_{j=1}^{n}x_{\mathrm{d},j}$. 
Thus when using zones of twice the thickness, i.e. $x_{\mathrm{d},j}=2h$, one obtains
\begin{equation}
A = A_{\mathrm{2h}} + \varepsilon_{\mathrm{2h}} \quad ,
\label{EQA2h}
\end{equation}
and similar for further refinement or coarsening.

Taylor-series analysis of truncation errors suggests that the error $\varepsilon_h$ is proportional to some power $p$ of the zone thickness $h$, i.e. $\varepsilon_h \propto h^p$. It follows that the error with  a twice coarser spacing is
\begin{equation}
\varepsilon_{\mathrm{2h}} = 2^{p} \varepsilon_{\mathrm{h}}\quad ,
\label{EQehconvergence}
\end{equation}
where $p$ is the order of convergence. Setting Eqs. (\ref{EQAh}) and (\ref{EQA2h}) as equal and inserting Eq. (\ref{EQehconvergence}) leads to 
\begin{equation}
\varepsilon_{\mathrm{h}} = \frac{A_{\mathrm{h}}-A_{\mathrm{2h}}}{2^{p}-1} \quad .
\label{EQeh}
\end{equation}

Insert Eq. (\ref{EQehconvergence}) into Eq. (\ref{EQeh}) written for $\varepsilon_{\mathrm{2h}}$ finally gives
\begin{equation}
p = \frac{\log \left( \frac{A_{\mathrm{2h}}-A_{\mathrm{4h}}}{A_{\mathrm{h}}-A_{\mathrm{2h}}} \right)}{\log \left( 2 \right) } \quad .
\label{EQp}
\end{equation}

Figures \ref{FIGCr0gdisc} and \ref{FIGCr0gdiscLOG} show that the deviation of $C_{\mathrm{R}}$ can differ depending on $\gamma$. Thus to estimate $\varepsilon_{\mathrm{h}}$ and $p$ in Eqs. (\ref{EQeh}) and (\ref{EQp}), set
\begin{equation}
A_{\mathrm{2h}} - A_{\mathrm{4h}} \approx \max \{ C_{\mathrm{R,2h}}(\gamma) - C_{\mathrm{R,4h}}(\gamma) \} \quad ,
\label{EQA2hA4h}
\end{equation}
\begin{equation}
A_{\mathrm{h}} - A_{\mathrm{2h}} \approx \max \{ C_{\mathrm{R,h}}(\gamma) - C_{\mathrm{R,2h}}(\gamma) \} \quad ,
\label{EQAhAh}
\end{equation}
where $C_{\mathrm{R,h}}(\gamma)$ is the reflection coefficient for zone thickness $x_{\mathrm{d},j}=h$ and forcing strength $\gamma$, and  the $\max \{X(\gamma)\}$-function delivers the maximum value of $X$ of all values $\gamma$ in the considered range $0.625\, \mathrm{rad/s} \leq \gamma \leq 10\, 000\, \mathrm{rad/s}$.

\begin{figure}[H]
\includegraphics[width=\linewidth]{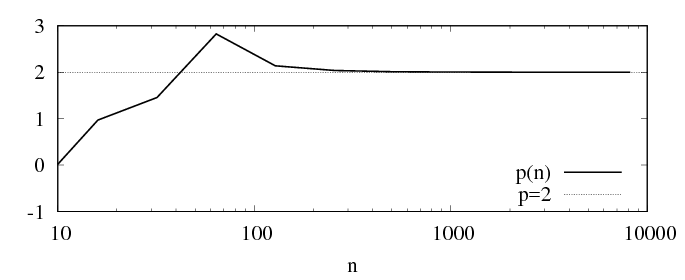} 
\caption{Order of convergence $p$ over number of zones $n$, for forcing as in Figs. \ref{FIGCr0gdisc} and \ref{FIGCr0gdiscLOG}} \label{FIGcompareDisc}
\end{figure}

Figure \ref{FIGcompareDisc} shows, exemplarily for $b(x)$ according to Eq. (\ref{EQblendexp}),  that the analytical solution converges with $2^{\mathrm{nd}}$ order to the analytical solution for $n=\infty$.  For the blending functions investigated in Appendix B, all curves showed $2^{\mathrm{nd}}$ order convergence (i.e. $p \rightarrow 2$ if $n \rightarrow \infty$), except constant blending ($b(x)=1$), for which the solution naturally must be exact independent of the number of zones. 
 It is out of the scope of this work to rigorously prove that for all possible $b(x)$ the order of convergence will be at least $p=2$, so this is left for future research; however, the present results suggest that this is the case.

\begin{figure}[H]
\includegraphics[width=\linewidth]{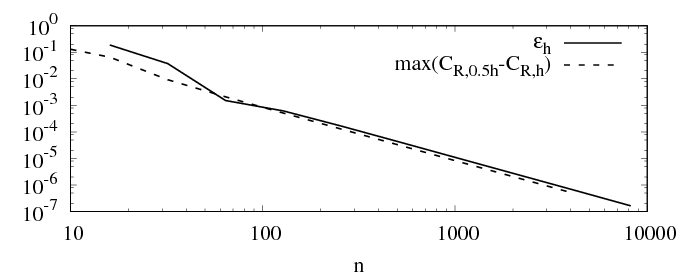} 
\caption{Error estimate $\varepsilon_{\mathrm{h}}$  over number of zones $n$, for forcing as in Figs. \ref{FIGCr0gdisc} and \ref{FIGCr0gdiscLOG}; $\max \{ C_{\mathrm{R,0.5h}} - C_{\mathrm{R,h}} \}$ gives the largest difference between reflection coefficients $C_{\mathrm{R}}$ for $n$ and $2n$ zones} \label{FIGcompareDisc2}
\end{figure}

Figure \ref{FIGcompareDisc2} shows that the error estimate $\varepsilon_{\mathrm{h}}$ for different $n$ decays accordingly, and the difference  between greatest discrepancies of reflection coefficients for zone thicknesses $h$ and $0.5h$ shows the same rate of decay and lies below  $\varepsilon_{\mathrm{h}}$. Thus for practical grids in flow simulations, as well as for the theory results plotted in this work, the forcing zone performance can be assumed independent of the number $n$ of zones or grid cells.

Given the simulation setup in Sect. \ref{SECsetup}, it is expected that, when the grid resolution increases, the wave damping behavior in the flow simulation will converge towards the solution for the specified continuous blending function $b(x)$.
Since Sect. \ref{SECresdisc} showed that the theory from Sect. \ref{SECtheory} predicts flow simulation results with great accuracy, it is expected that the results of the theory from Sect. \ref{SECtheory}, which is based on discontinuous piece-wise constant blending $b(x)$, will converge towards the analytical solution for any given continuous blending function  $b(x)$, if the forcing zone is subdivided into a sufficient number of zones $n$.

\section*{Acknowledgements}
The study was supported by the Deutsche Forschungsgemeinschaft (DFG).



\end{document}